\documentclass[twocolumn]{revtex4-2}

\usepackage{graphicx}
\usepackage{amsmath,amssymb}
\usepackage{color}
\usepackage{hyperref}
\usepackage{placeins}
\usepackage{float}
\usepackage{appendix}
\usepackage{multirow}
\usepackage{soul}
\usepackage{booktabs} % For better looking tables

\begin{document}

\title{Thermo-coalescence model for Light Nuclei production in Relativistic Heavy-Ion Collisions}

\author{Arun Kumar Yadav}
\email{ak.yadav@vecc.gov.in}
\affiliation{Variable Energy Cyclotron Centre, HBNI, 1/AF Bidhan Nagar, Kolkata 700 064, India~}
\affiliation{Homi Bhabha National Institute, Anushakti Nagar, Mumbai  400094, India}

\author{Nachiketa Sarkar}
\email{nachiketa.sarkar@gmail.com }
\affiliation{Department of Physical Sciences, Indian Institute of Science Education and Research Berhampur, Laudigam - 760003, Ganjam, Odisha, India}
\affiliation{Homi Bhabha National Institute, Anushakti Nagar, Mumbai 400094, India}

\author{Sudhir Pandurang Rode}
\email{sudhirrode11@gmail.com }
\affiliation{Veksler and Baldin Laboratory of High Energy Physics, Joint Institute for Nuclear Research, Dubna - 141980, Moscow region, Russian Federation}

\author{Partha Pratim Bhaduri}
\email{Corresponding author:partha.bhaduri@vecc.gov.in }
\affiliation{Variable Energy Cyclotron Centre, HBNI, 1/AF Bidhan Nagar, Kolkata 700 064, India}
\affiliation{Homi Bhabha National Institute, Anushakti Nagar, Mumbai 400094, India}

\author{Abhijit Bhattacharyya}
\email{abhattacharyyacu@gmail.com}
\affiliation{Department of Physics, University of Calcutta, 92, A. P. C. Road, Kolkata - 700009, India }

\author{Amaresh Jaiswal}
\email{jaiswal.amaresh@gmail.com}
\affiliation{School of Physical Sciences, National Institute of Science Education and Research, Jatni - 752050, Odisha, India }
\affiliation{Homi Bhabha National Institute, Anushakti Nagar, Mumbai 400094, India}

\date{\today}

\begin{abstract}
  We employ a hybrid approach to describe the light nuclei production mechanism where the nucleons are assumed to be thermally produced, and are allowed to form light nuclei using a coalescence prescription. In this approach, we first fit transverse momentum ($p_{T}$) distribution of nucleons using hydro-inspired boost-invariant blast-wave model. The extracted parameters are then used to describe the deuteron $p_{T}$ spectra, along with two additional parameters that characterize the coalescence prescription employed in this study. We refer this combined approach as ``thermo-coalescence model'' and it is designed to study the deuteron production and describe the experimental measurements. In this work, we analyze the measured $p_{T}$ distribution of protons and deuterons from Pb-Pb collisions at the ALICE Collaboration at LHC. We also evaluate the $p_{T}$-integrated deuteron yields using this approach and compare with experimental measurements. A Bayesian inference framework is employed to determine the best-fit parameters of the thermo-coalescence model. Finally, we estimate the traditionally used experimental coalescence parameter ($B_{A}$) within our framework in order to establish a connection between our model and the conventional coalescence approach commonly used to relate experimental data with theoretical descriptions of light nuclei production. 
\end{abstract}
\maketitle
\section{Introduction}
\label{intro}

The observation of light (anti-)nuclei in high-energy heavy-ion collisions has garnered significant attention from the global heavy-ion collision community, with experiments conducted at facilities ranging from the Large Hadron Collider (LHC)~\cite{ALICE:2017nuf, ALICE:2017jmf} and the Relativistic Heavy Ion Collider (RHIC)~\cite{STAR:2023uxk, Liu:2022ump, STAR:2021ozh} to those at Super Proton Synchrotron (SPS) energies~\cite{Kolesnikov:2015ola, NA49:2011blr, NA49:2016qvu}. Relativistic heavy-ion collisions offer a unique environment to investigate the formation mechanisms of light nuclei. At LHC energies, the baryonic chemical potential ($\mu_{B}$)  is nearly zero, resulting in the production of anti-protons and anti-neutrons in quantities comparable to their matter counterparts. In contrast, at RHIC energies, $\mu_{B}$ remains on the order of a few MeV. Despite these $\mu_{B}$ values, significant production of light anti-nuclei have been measured at RHIC, leading to the first observation of the anti-alpha~\cite{STAR:2011eej} and anti-hypertriton~\cite{STAR:2010gyg} by the STAR collaboration.

The production of light (anti-)nuclei has been studied experimentally across a wide range of energies and facilities around the world. Beginning with early observations at Bevalac~\cite{Nagamiya:1981sd}, such studies have extended to the Schwerionensynchrotron (SIS)~\cite{FOPI:2010xrt, HADES:2020lob}, the Alternating Gradient Synchrotron (AGS)~\cite{E886:1994ioj, E864:2000auv, E864:2000loc, Albergo:2002gi, E878:1998vna}, the Super Proton Synchrotron (SPS)~\cite{NA49:2000kgx, NA49:2004mrq, NA44:1995lds, NA52:2003udo}, the Relativistic Heavy Ion Collider (RHIC)~\cite{STAR:2011eej, STAR:2001pbk, STAR:2019sjh, STAR:2009kaf, STAR:2021ozh, STAR:2020hya, PHENIX:2004vqi, PHENIX:2007tef}, and the Large Hadron Collider (LHC)~\cite{ALICE:2022zuz, ALICE:2022xiu, ALICE:2017nuf, ALICE:2015rey, ALICE:2015wav}, as well as in smaller collision systems~\cite{ALICE:2015wav}. A key puzzle arises from the observation that light nuclei—despite being very loosely bound, with binding energies of the order of few MeV per nucleon—are produced in environments with temperatures on the order of 100–160 MeV. This apparent paradox has made the understanding of their production mechanisms a subject of ongoing debate. Two main theoretical approaches have been developed to explain this phenomenon: thermal models and coalescence-based models.

%The latter observation has forced physicists to add third axis into positive strangeness direction and now, there is hope for more discoveries in this direction in future. Light (anti)-nuclei production has been studied experimentally are various energies and experiments across different facilities around the globe. Starting from Bevalac~\cite{Nagamiya:1981sd}, the Schwerionensynchrotron (SIS)~\cite{FOPI:2010xrt, HADES:2020lob}, the Alternating Gradient Synchrotron (AGS)~\cite{E886:1994ioj, E864:2000auv, E864:2000loc, Albergo:2002gi, E878:1998vna}, the SPS~\cite{NA49:2000kgx, NA49:2004mrq, NA44:1995lds, NA52:2003udo}, the RHIC~\cite{STAR:2011eej, STAR:2001pbk, STAR:2019sjh, STAR:2009kaf, STAR:2021ozh, STAR:2020hya, PHENIX:2004vqi, PHENIX:2007tef}, and the LHC~\cite{ALICE:2022zuz, ALICE:2022xiu, ALICE:2017nuf, ALICE:2015rey, ALICE:2015wav}, and in smaller collision systems~\cite{ALICE:2015wav}. A key question arises from the fact that light nuclei are very loosely bound—binding energies range from a few keV to about 8 MeV per nucleon—yet are produced in an environment with temperatures on the order of 100–160 MeV. This apparent paradox has made their production mechanisms an area of active debate. Two broad theoretical frameworks aim to explain their formation: thermal models and coalescence-based models.

The statistical approach, namely thermal models, has been remarkably successful in describing hadron yields. These models have also been applied for decades to describe the production of light nuclei~\cite{Mekjian:1977ei, Gosset:1978pqf, Mekjian:1978us, Siemens:1979dz, Stoecker:1981za, Hahn:1986mb, Csernai:1986qf}. Notably, studies such as~\cite{Donigus:2020ctf, Donigus:2020fon} have shown that the inclusion or exclusion of light nuclei yields in thermal model fits can lead to significant differences in the extracted chemical freeze-out temperature ($T_{ch}$) and baryon chemical potential ($\mu_{B}$), particularly in the finite-$\mu_{B}$ region corresponding to low beam energies on the QCD phase diagram. A major point of contention, however, lies in the fact that the extracted $T_{ch}$ is around 156 MeV which is significantly higher than the binding energies of light nuclei (e.g., 2.2 MeV for deuterons). Despite this, it has been observed at RHIC and LHC energies that even when only light nuclei yields are used to determine the freeze-out temperature, the resulting value remains close to $T_{ch}\approx$ 160 MeV~\cite{Andronic:2017pug, Braun-Munzinger:2018hat, ALICE:2017jmf, Biswas:2020kpu}. This apparent contradiction, where loosely bound objects are seemingly produced in a hot medium, has been famously referred to as the ``snowballs in hell" problem~\cite{Braun-Munzinger:1995uec, Mrowczynski:2020ugu, Oliinychenko:2018ugs}.

%The statistical approach, i.e. thermal models, have been extremely successful in interpreting the hadron yields. Thermal models have been applied to describe the production yield of light nuclei for many years~\cite{Mekjian:1977ei,Gosset:1978pqf,Mekjian:1978us,Siemens:1979dz,Stoecker:1981za,Hahn:1986mb,Csernai:1986qf}. Interestingly, it has been found in~\cite{Donigus:2020ctf,Donigus:2020fon} that the inclusion or exclusion of nuclei production yields in the thermal model calculations to extract chemical freeze-out temperature ($T_{ch}$) and baryon chemical potential ($\mu_{B}$) show significant differences in the region of finite $\mu_{B}$ corresponding to low beam energies on the QCD phase diagram. However, the argument that does not serve in favor of this approach is that the value of $T_{ch}$ ($T_{ch}\approx$ 156 MeV) is much higher than the binding energy of the light nuclei, for instance 2.2 MeV for deuteron. Nevertheless, at RHIC and LHC, it is seen that even if only light nuclei are used for the extraction of the temperature, the result is quite close to 156 MeV, which is $T_{ch}\approx$ 160 MeV~\cite{Andronic:2017pug, Braun-Munzinger:2018hat, ALICE:2017jmf}. This contradictory picture where the objects with small binding energy produce at 160 MeV lead to an argument of ``snowballs in hell''~\cite{Braun-Munzinger:1995uec, Mrowczynski:2020ugu, Oliinychenko:2018ugs}.

On the other hand, the coalescence model, which is a microscopic approach to light nuclei production, was first introduced in the 1960s to interpret data from the proton synchrotron at CERN, where various targets were bombarded with a 25 GeV proton beam for the first time~\cite{Cocconi:1960zz}. This mechanism, in which deuterons are formed by the coalescence of protons and neutrons in close proximity in phase space, was proposed to explain the unexpectedly large deuteron cross-section observed in proton–nucleus collisions~\cite{Butler:1961pr, Butler:1963pp, Schwarzschild:1963zz}. Since then, the coalescence model has undergone significant development, with numerous efforts to describe cluster yields in heavy-ion collisions across a wide range of energies. The first applications in this context began in the 1970s at the Bevalac at Lawrence Berkeley Laboratory~\cite{Gutbrod:1976zzr, Gosset:1976cy, EOS:1994jzn, Nagamiya:1981sd}. Subsequent implementations were carried out in experiments such as E802/E866, E814, E864, E877, and E878 at the Alternating Gradient Synchrotron (AGS) at Brookhaven National Laboratory (BNL)~\cite{E886:1994ioj, E-802:1994hea, E814:1994kon}, and in NA44, NA49, and NA52 at CERN for interpreting heavy-ion data~\cite{NA44:1999uvg, NA49:2011blr, NA52NEWMASS:1996ptr, NA52:2003udo}.

The coalescence model also proved effective in describing light nuclei yields at RHIC~\cite{STAR:2009kaf, PHENIX:2007tef, BRAHMS:2010nkj}. More recently, it has been incorporated into modern phenomenological simulations based on transport models such as \texttt{SMASH}, \texttt{UrQMD}, \texttt{AMPT}, and \texttt{THESEUS}, providing theoretical predictions consistent with experimental observations~\cite{Bailung:2024sca, Bailung:2023dpv, Reichert:2023tww, Kozhevnikova:2022wms}. Additionally, methods such as the Minimum Spanning Tree (MST) algorithm have been developed to identify nucleon and cluster configurations within dynamical simulations~\cite{Aichelin:1991xy, Aichelin:2019tnk}, and have been implemented in models like \texttt{PHQMD}~\cite{Kireyeu:2022qmv} to identify clusters more reliably. The coalescence approach has yielded several promising predictions and interpretations. Notably, in analogy with the observed quark-number scaling of hadronic elliptic flow, a mass-number scaling of elliptic flow for light nuclei has also been explored with some degree of success~\cite{STAR:2016ydv,ALICE:2019ikx,ALICE:2020chv,ALICE:2024say}, lending strong support to coalescence and recombination based models.

Despite their widespread use, both thermal and traditional coalescence models exhibit significant limitations in explaining light nuclei production in high-energy heavy-ion collisions. The thermal model, although highly successful in reproducing hadron yields, assumes chemical freeze-out temperatures in the range of approximately 156–160 MeV which is substantially higher than the binding energies of light nuclei such as deuterons ($\simeq$2.2 MeV). This disparity raises concerns about the model’s applicability to loosely bound nuclear states. Furthermore, the thermal approach lacks a dynamical treatment of the formation process and does not incorporate the space-time evolution crucial for realistic cluster formation. In contrast, conventional coalescence models adopt simplified geometric criteria based on nucleon proximity in phase space and often rely on empirical parameter tuning to match observed yields. While these models can effectively describe the momentum spectra of light nuclei, they frequently struggle to reproduce integrated yield ratios consistently across varying collision energies and centralities.

%Despite their widespread use, both thermal and traditional coalescence models exhibit notable limitations in explaining light nuclei production in high-energy heavy-ion collisions. The thermal model, while successful in describing hadron yields, assumes chemical freeze-out temperatures (~156–160 MeV) that far exceed the binding energies of light nuclei, raising the “snowballs in hell” paradox and casting doubt on its applicability to loosely bound states like deuterons. Moreover, it lacks dynamical formation mechanisms and cannot capture the space-time evolution relevant for cluster formation. On the other hand, conventional coalescence models rely on simplified geometrical criteria for nucleon proximity in phase space and often require empirical tuning to reproduce absolute yields. These models tend to perform well in describing momentum spectra but struggle to simultaneously account for integrated yield ratios across different energies and centralities. On experimental front, a significant progress in realizing the production mechanism has been made, though, still no clear picture has emerged so far.

%---------------------------------------------------------------------
\begin{figure}[h!]
\includegraphics[scale=0.48]{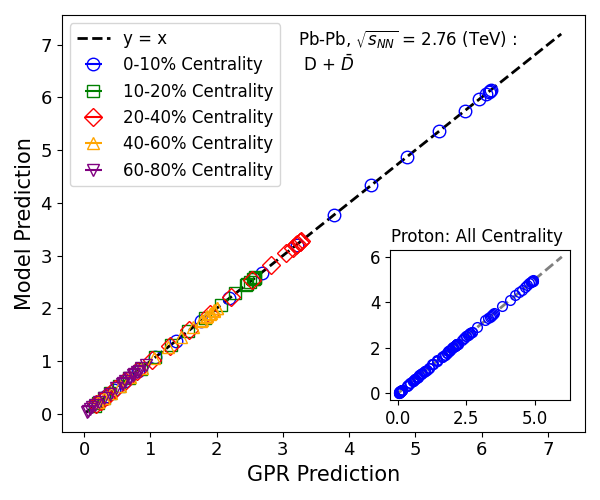}
\includegraphics[scale=0.48]{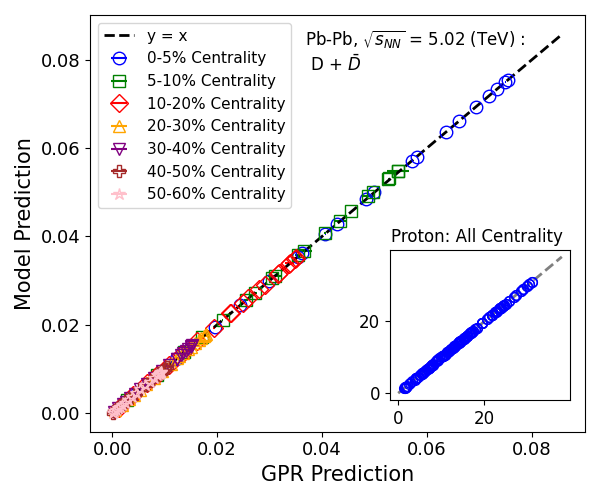}
    
\caption{Validation of Gaussian Process Regression (GPR) emulators for deuteron spectra at two different collision energies: $\sqrt{s_{NN}} = 2.76$ TeV (left panel) and $\sqrt{s_{NN}} = 5.02$ TeV (right panel). Different colored markers represent the invariant yield at varying transverse momentum ($p_T$) values, calculated using 10 arbitrary parameter sets for each centrality class obtained from both explicit model calculations and emulator predictions. The dashed black line represents the ideal $y = x$ relationship, indicating perfect agreement between the emulator and model. The inset plots present similar comparisons for protons across all centrality classes.}
\label{fig:GPR_Validation}
\end{figure}

%\begin{figure*}[ht]
%\begin{center}
%\includegraphics[scale=0.5]{Figures/Sobol_2_76.png}
%\includegraphics[scale=0.5]{Figures/Sobol_5_02.png}
%\end{center}
%\caption{Sobol sensitivity indices for deuteron and proton $p_T$ spectra across different centrality classes in Pb+Pb collisions measured by ALICE collaboration at LHC at $\sqrt{s_{NN}} = 2.76$ TeV (top panel) and at $\sqrt{s_{NN}} = 5.02$ TeV (bottom panel). Each grid cell represents a specific $p_T$ value, arranged from low to high along the horizontal axis. The right panel illustrates the sensitivity of coalescence model parameters $\alpha$ and $C_D$ for deuteron spectra, while the left panel shows the sensitivity of thermal blast wave parameters for proton spectra. Color bars represent the total Sobol index (ST) values, and red horizontal lines demarcate the proton centrality bins.}
	%
%\label{fig2:Sobol_index}
%\end{figure*}

\begin{figure*}[ht]
\begin{center}
\includegraphics[scale=0.5]{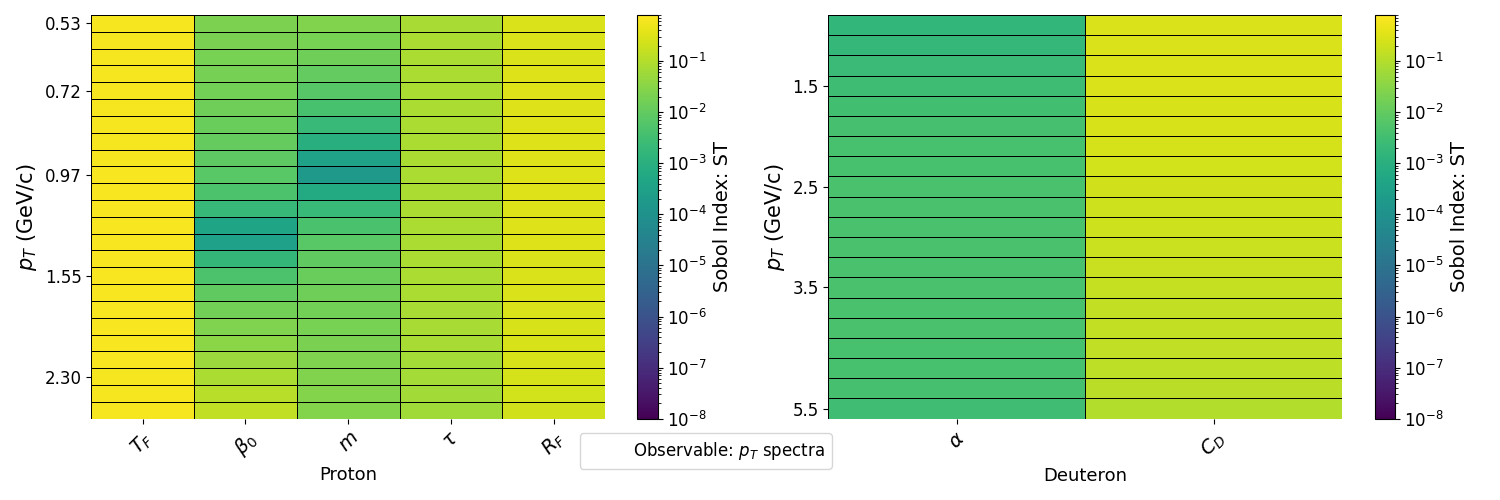}
 \end{center}
\caption{Sobol sensitivity indices for the transverse momentum ($p_T$) spectra of protons and deuterons in Pb-Pb collisions are presented. Each grid cell corresponds to a specific $p_T$ bin, top to bottom along the vertical the axis from low to high $p_T$. The parameter ranges used in this analysis were chosen to encompass the full range of values obtained from Bayesian fits across all centrality classes and collision energies considered in this study. The left panel illustrates the sensitivity of the proton $p_T$ spectra to thermal blast-wave model parameters, while the right panel shows the sensitivity of the deuteron spectra to the coalescence model parameters $\alpha$ and $C_D$. Color bars indicate the total Sobol index ($S_T$) values, which quantify the contribution of each parameter to the variance of the corresponding observable.
}
\label{fig2:Sobol_index}
\end{figure*}

\begin{figure*}[ht]
    \centering
    % First row
 
    \includegraphics[width=0.495\textwidth]{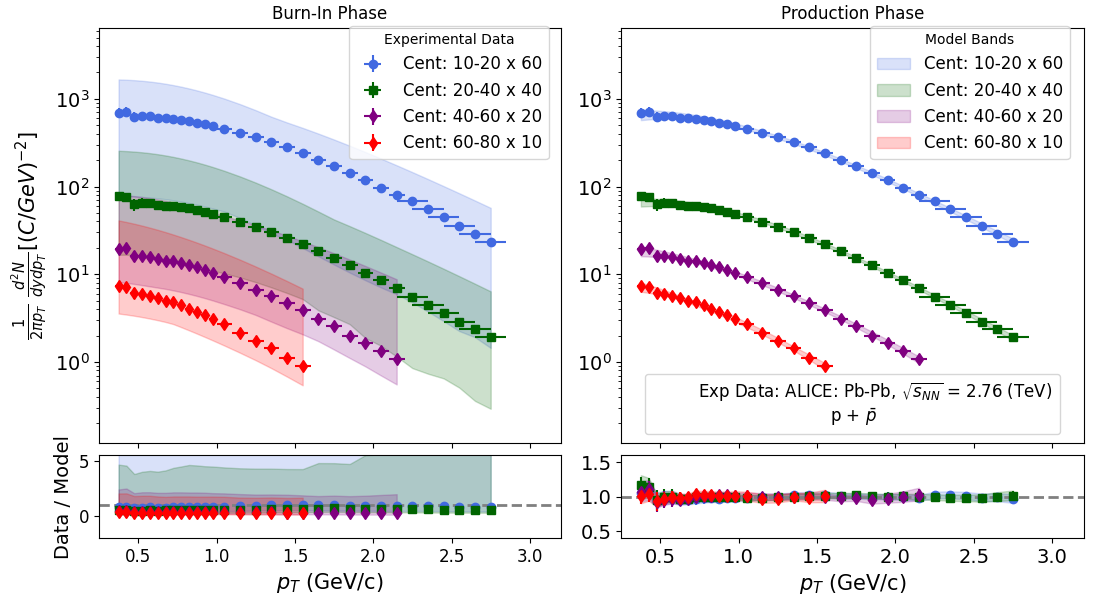}
    \includegraphics[width=0.495\textwidth]{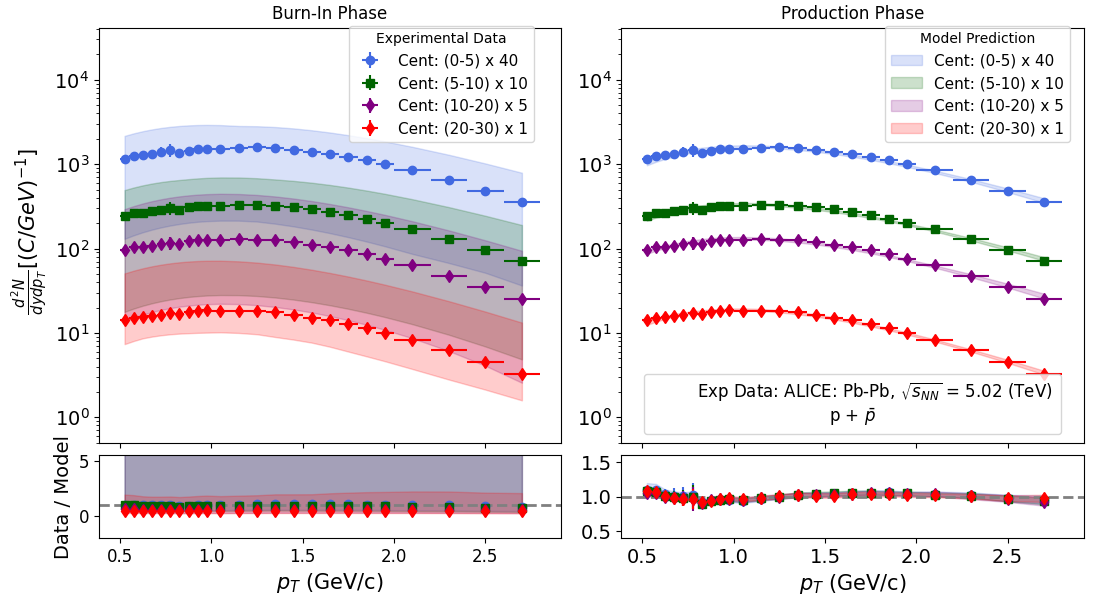}
    \\[2mm] % Small vertical space between rows

    % Second row
    \includegraphics[width=0.495\textwidth]{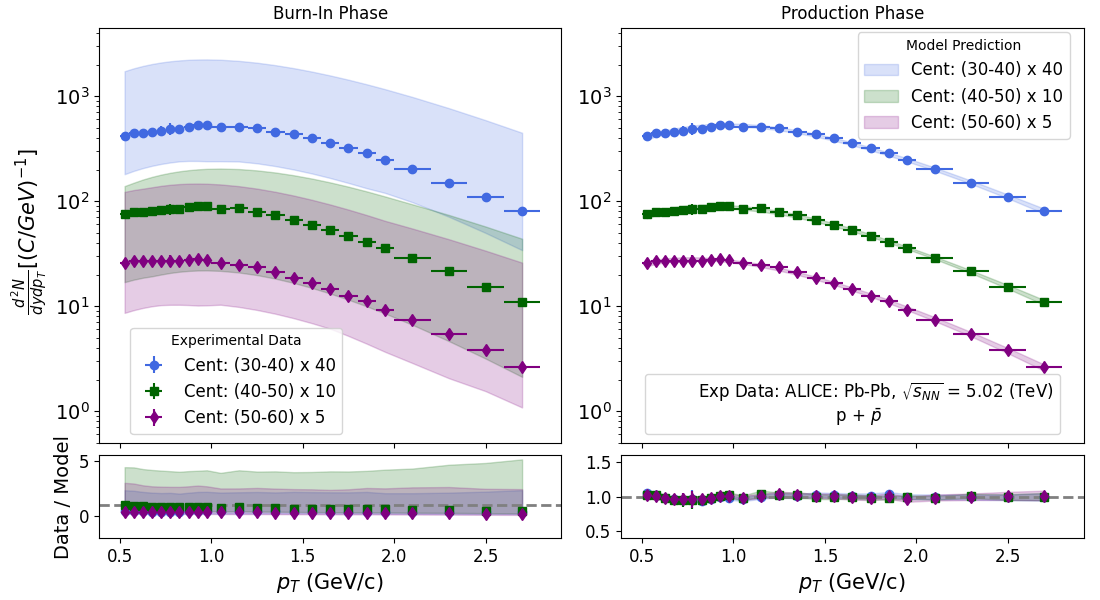}
   
  \caption{Comparison of proton spectra from model predictions and experimental measurements from ALICE at the LHC in Pb-Pb collisions at $\sqrt{s_{NN}} = 2.76$ TeV (Top Left) and $\sqrt{s_{NN}} = 5.02$ TeV (Top Right and Bottom) across different centrality classes. Experimental data points (shown with various markers) include error bars, while model predictions are depicted as shaded bands. The top panels display emulator predictions during the Burn-In phase (sampling from the prior distribution) and the bottom panel shows predictions based on posterior sampling.}

    \label{fig3:exp_model_com_proton}
    
\end{figure*}

In this work, we employ a unified framework that combines thermal and coalescence mechanisms, a ``thermo-coalescence model" to describe light nuclei production in relativistic nuclear collisions\cite{Florkowski:2023uim,Drogosz:2025vdq}. The central assumption is that nucleons are thermally produced and can be described using a thermal model, in our case boost-invariant blast-wave model. Thermally produced nucleons that are sufficiently close in phase space are then assumed to undergo coalescence to form light nuclei. We apply this hybrid approach to analyze the transverse momentum ($p_{T}$) spectra of protons and deuterons in Pb–Pb collisions measured by the ALICE Collaboration at the LHC. One major advantage of ALICE detector setup is its excellent particle identification and vertexing capabilities. Light nuclei measurement is performed using data recorded by the Inner Tracking System (ITS), the Time Projection Chamber (TPC) and the Time-Of-Flight (TOF) detector, all located inside a homogeneous magnetic field of strength of 0.5 T and cover the full azimuthal acceptance and the pseudo-rapidity range, $|\eta| < 0.9$. While TPC provides particle identification via the specific energy loss ($dE/dx$), TOF identifies the light (anti)-nuclei by determining their velocity. A combined analysis of TPC and TOF data, enables the identification of (anti)deuteron spectra up to $p_{T} = 6$ GeV/$c$ in Pb-Pb collisions. At $\sqrt{s_{NN}} = 2.76$ TeV using data from High Momentum Particle Identification Detector (HMPID), the production spectra has been measured to $p_{T} =8$ GeV/$c$. 
Since our study is limited to mid-rapidity region at LHC energies, the assumption of longitudinal boost-invariance remains valid, justifying the use of the boost-invariant version of the blast-wave model. We adopt a Bayesian inference framework to extract the best-fit parameters of our thermo-coalescence model. The parameters extracted from fits to the proton spectra are subsequently used to compute the deuteron distributions through a coalescence formalism inspired by Ref.~\cite{Fries:2008hs}.

%\textbf{\textit{There have been various attempts to combine these two approaches under single framework to describe the nuclei production~\cite{Wang:2023rpd,Wang:2024jpe,Yin:2017qhg}. Though, there are similarities in the types of model that are employed, the treatment and details in our study are different. For instance, we have used a coalescence model which was first used to coalesce quarks into hadrons~\cite{Fries:2003kq,kuiper}. For the purpose of this study, this approached is adjusted to perform the nucleon coalescence. In the blast-wave model, all five parameters are kept free, and such challenging fits are performed using Bayesian inference which makes the parameters more credible and precise. [\color{red} more differences?]}}

In literature there have been previous attempts to unify these two approaches within a single framework to describe light nuclei production~\cite{Yin:2017qhg, Zhu:2017zlb, Wang:2023rpd, Wang:2024jpe} in nuclear collisions at RHIC and LHC. In~\cite{Zhu:2017zlb} the authors have employed a coalescence model, based on the phase-space distribution of nucleons from an extended blast-wave model that allows high $p_{T}$ nucleons  to be more spread in space when their momenta are more aligned along the reaction plane, to describe the spectra and elliptic flow ($v_{2}$) of deuteron and helium-3 in $\sqrt{s_{NN}} = 2.76$ TeV Pb-Pb collisions. On the other hand in~\cite{Wang:2023rpd, Wang:2024jpe}, the authors presented an improved nucleon coalescence framework by including the coordinate-momentum correlation in nucleon joint distribution to describe light nuclei production in Pb-Pb collisions at LHC. The momentum distribution of primordial nucleons, in this approach were obtained by performing a blast wave model fit to the measured prompt proton $p_{T}$ spectra.  While there are similarities in the classes of models employed, the methodology and specific implementations in our work differ significantly. The coalescence process has been formulated in a covariant framework using Wigner functions, specifically to describe the coalescence of baryons into light nuclei and clusters in nuclear collisions as well as quark-to-hadron coalescence~\cite{Dover:1991zn, Scheibl:1998tk, Fries:2003kq, kuiper}. We employ this model, together with the blast-wave model for nucleon spectra, to investigate light nucleus production through nucleon coalescence. No space-momentum correlation has been assumed either in phase space distribution of the nucleons or in their joint distribution during fusion. Additionally, in our adopted blast-wave model, all five parameters are treated as free and are fitted using Bayesian inference. This not only allows for a more robust handling of uncertainties but also enhances the precision and reliability of the extracted parameters.

%In this work, we introduce a combined approach of thermal and coalescence mechanism, a ``thermo-coalescence model''  to describe the light nuclei production. The assumption is that the nucleons are thermally produced and therefore, they are described using a thermal model, in our case boost-invariant blast-wave model. The thermally produced nucleons, which are close in phase space undergo coalescence to form light (anti)nuclei. We have applied this combined approach to analyze the transverse momentum distributions of protons and deuterons in Pb-Pb collisions at the ALICE experiment at LHC. It is worth  mentioning that since we are only restricting our work to available LHC energies, the assumption of boost-invariance is still intact and therefore the model is suitable in boost-invariant mode. The parameters that obtained from the fits are then used to describe the light nuclei distribution using coalescence approach inspired from Ref.~\cite{Fries:2008hs}. 

This rest of the article is divided into following sections. Section \ref{Blast_wave fit of hadrons} and Section \ref{two nucleon coalescence} describe the blast-wave and coalescence approach that are used in this work, respectively. In Section \ref{results and discussions}, the obtained results are discussed and interpreted. Finally, we conclude in Section \ref{conclusions}.

\section{Blast wave fit of light hadrons}
\label{Blast_wave fit of hadrons}

Blast-wave model is a hydro-inspired phenomenological model widely used in heavy-ion collisions to describe the $p_{T}$ distributions of the various hadron species emitted at the freeze-out of the fireball. In this work, a boost-invariant blast-wave model is employed, details of which can be found here~\cite{Schnedermann:1993ws}. Here, we briefly introduce the main features of the model for completeness. Within this boosted thermal phenomenology, the single particle momentum distribution following Cooper-Frye formalism for modeling kinetic freeze-out can be written as:

\begin{equation}
\label{CF}
E\frac{d^3N}{d^3p} = {\frac{g}{2 \pi^{3}}}\int d^{3}\Sigma_{\mu}^{f}(x)p^{\mu}f_{eq}(x,p)
\end{equation}

where $g$ denotes the degeneracy factor. In the model, $f_{eq}(x,p)$ is assumed to be the equilibrium distribution function.  The freeze-out hyper surface $\Sigma^f_\mu(x)$ is evaluated from freeze out criterium for thermal decoupling. 

For an expanding fireball under local thermal equilibrium, the Fermi-Dirac distribution for nucleons, is given by:
%
\iffalse
For an expanding fireball under local thermal equilibrium, the boosted thermal distribution is given by:
\begin{equation}
\label{dist}
f_{eq}(x,p) = \exp\left( -{\frac{p.u(x) - \mu(x)}{T(x)}} \right)
\end{equation}
\fi
%
\begin{equation}
\label{dist}
f_{eq}(x,p) = \frac{g}{\exp[(u^{\mu} p_{\mu})/T] + 1},
\end{equation} 
where $g$ is the degeneracy factor.% and $P=(E,x{\bf P})$.

We consider irrotational and boost invariant flow. For boost invariant flow, without loss of generality, we can consider $z=0$ slice such that $v^z=z/t=0$. Performing fit to the $p_{T}$ spectra of protons should then fix the hydrodynamic quantities like temperature, $T$ and fluid velocity, $u^\mu=(u^\tau,\,u^r,\,u^\phi,\,u^z)=\gamma_T(1,\,\beta_T,\,0,\,0)$ as:
\begin{align}
T &= T_F , \label{BW1}\\
u^r &= \gamma_T \, \beta_T, \label{BW2}\\
u^\phi &= u^z= 0, \label{BW3}\\
u^\tau &= \gamma_T, \label{BW4}
\end{align}

Using the Milne co-ordinate system, the single particle transverse momentum spectra for a particle at the space-time point $(\tau, \eta_s ,r, \phi)$ with the four momentum $p^\mu$ = $(E, p_x , p_y , p_z )$ = $( m_T cosh y, p_T cos \phi_p , p_T sin \phi_p , m_T sinh y)$ can be written as~\cite{Schnedermann:1993ws}
\iffalse
\begin{align}
\label{therm}
\frac{d^2N}{d^2p_{T} dy} \propto \int_0^{R_{F}}r_\perp dr_\perp\ \int_0^{2\pi}d\phi \int_{-\infty}^{+\infty} \tau d\eta_s  m_T    cosh(y-\eta_s) f_{eq}
 \end{align}
\fi
\begin{multline}
\frac{d^2N}{d^2p_{T} dy} \propto \int_0^{R_{F}}r_\perp dr_\perp\ \int_0^{2\pi}d\phi \int_{-\infty}^{+\infty} \tau \, d\eta_s \,  m_T \,
\\
cosh(y-\eta_s) f_{eq} \, 
\label{therm}
\end{multline}

%We first fit proton spectra using blast wave model to obtain the kinetic freeze-out temperature and fluid flow velocity. 

where $\eta_{s}$ denotes the space-time rapidity, $R_{F}$ is the transverse radius of the fireball at the kinetic freeze-out, $\gamma_T=1/\sqrt{1-\beta_T^2}$ is the Lorentz factor in the transverse direction and $\beta_T$ is the transverse expansion velocity. For central collisions, a power-law relation for the spatial profile of the transverse velocity leads to
\begin{equation}\label{betaT}
\beta_T = \beta_0\left( \frac{r_{\perp}}{R_{F}} \right)^m,
\end{equation}
where $\beta_0$ is the maximum transverse velocity and $n$ is the exponent of flow profile. The transverse flow velocity is related to transverse rapidity, $\rho$, thorough, $\rho(r_{\perp})={\rm tanh}^{-1}(\beta_T)$. The average transverse velocity can be obtained using expression, $\langle \beta_{T} \rangle=\frac{2}{2+m} \beta_{0}$. We use Eq.~\eqref{therm} to fit the transverse momentum spectra of protons, and extract the parameters of the fit, which are, the kinetic freeze-out temperature ($T_F$), maximum transverse flow velocity($\beta_0$), exponent of the transverse velocity profile ($m$), the freeze-out radius ($R_F$) and the freeze-out time ($\tau$).

Once we have fixed the hydrodynamic quantities at kinetic freeze-out from blast wave fit of the proton spectra, we focus on two nucleon coalescence to study the light nuclei production.

\begin{figure*}[t]
	\includegraphics[scale=0.295]{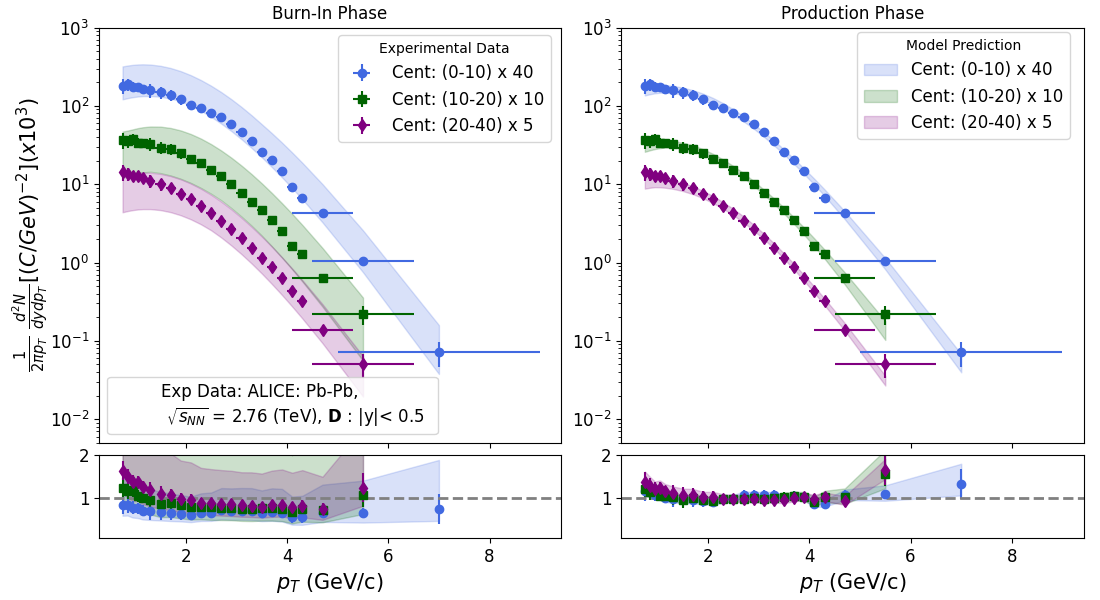}
	\includegraphics[scale=0.295]{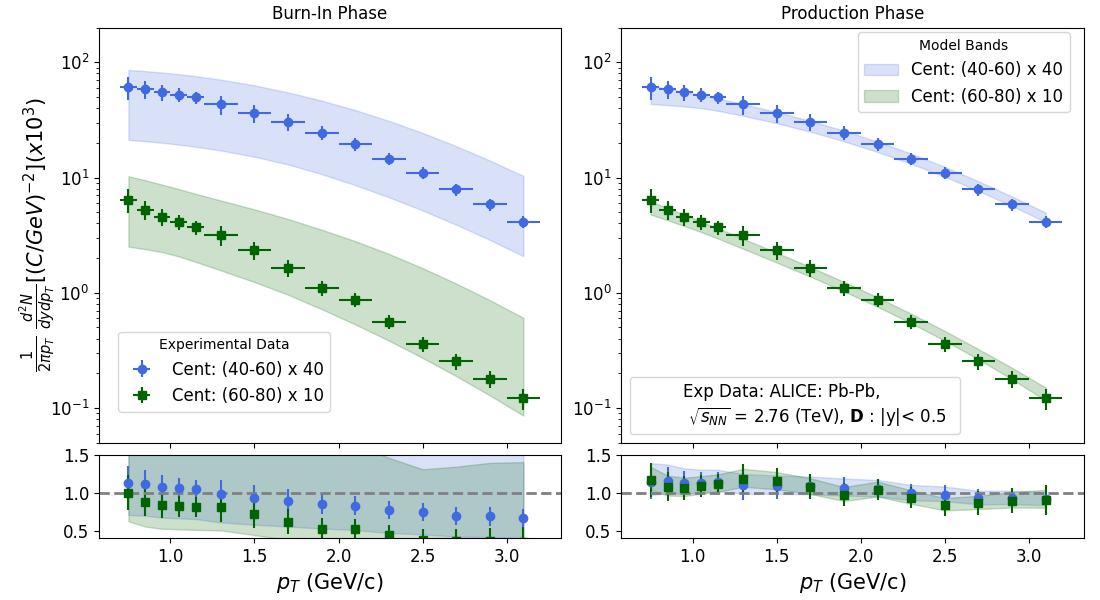}\\
	
	\includegraphics[scale=0.295]{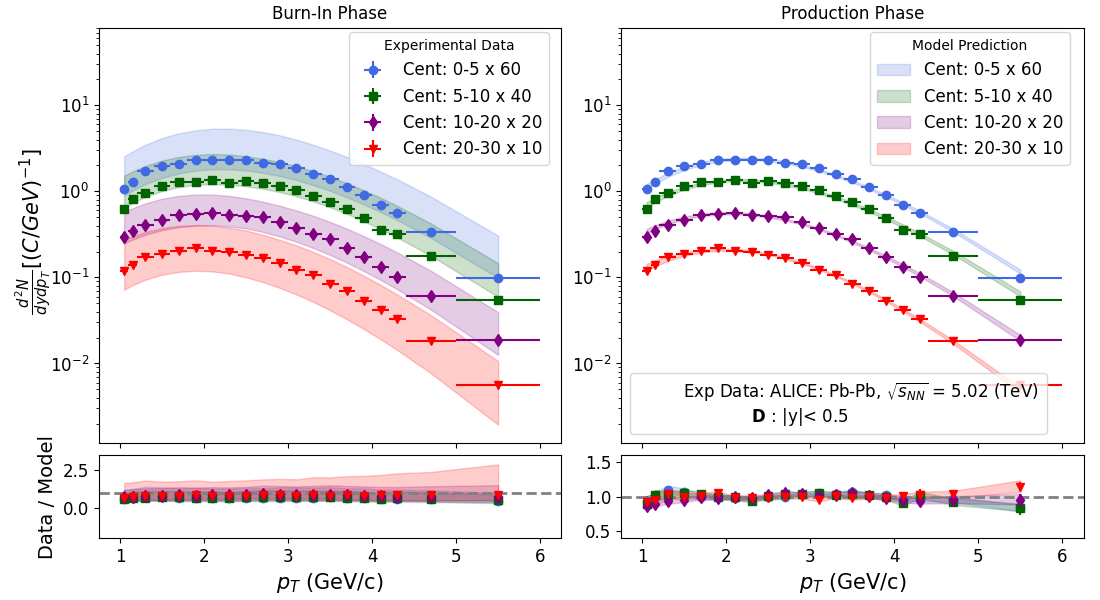}
	\includegraphics[scale=0.295]{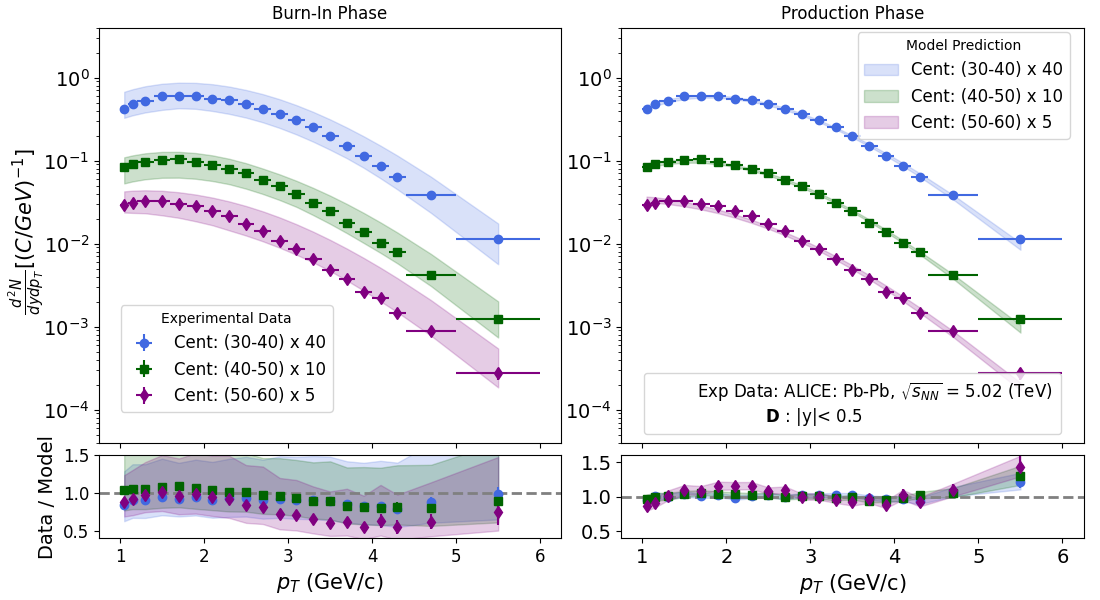}\\
	
	\caption{Same as the caption of Fig.~\ref{fig3:exp_model_com_proton}, but here the Bayesian fit is applied to deuteron spectra with two additional parameters: $\alpha$ and $C_D$. The thermal parameters are taken from the corresponding proton fit, with their values allowed to vary only within a narrow region around the respective median.}
	\label{fig1}
\end{figure*}

%---------------------------------------------------------------------

\section{Two nucleon coalescence}
\label{two nucleon coalescence}

The $p_{T}$ spectrum of light nuclei (in our case deuteron), formed by two nucleon coalescence at mid rapidity, can be  expressed as \cite{Fries:2003kq, kuiper}
\begin{align}
    \frac{dN_{\rm Deu}}{d^2 P_T dy}\Big|_{y=0} &= 
C_D \, \int_{\Sigma} d\Sigma_R \, \frac{P\cdot u(R)}{(2\pi)^3}\int_0^1 dx \nonumber\\
& f_{eq} \big(R,x\bm{P}\big) \left|\phi_D(x)\right|^2 f_{eq} \big(R,(1-x)\bm{P}\big),
\end{align}
where, $d\Sigma_R$ is the infinitesimal volume element of the hypersurface $\Sigma$, with $R\in\Sigma$ and $\phi_D(x)$ represents the wave function of the deuteron in the momentum space. In the above equation, $f_{eq}$ represents the phase-space distribution of the nucleons. Considering the phase-space distribution of the nucleons to be classical Maxwell-J\"uttner distribution and integrating over the space-time rapidity, one obtains~\cite{Fries:2003kq}
\begin{multline}
\frac{dN_{\rm Deu}}{d^2 P_T dy}\Big|_{y=0} = 
C_D \, \int_0^{R_F} r_\perp dr_\perp\, M_T \, I_0 \left[ \frac{P_T \sinh \rho(r_\perp)}{T_F}\right] 
\\
\int\limits_0^1 dx \> \left|\phi_D(x)\right|^2  k_D(x,P_T) \, ,
\label{deu_spec}
\end{multline}

Here, we use the notations

\begin{equation}
\begin{split}
  k_D(x,P_T) =  
  K_1 \Bigg[ \frac{\cosh \rho(r_\perp)}{T_F}
  \Big(& \sqrt{m_N^2 + x^2 P_T^2} \\
      & + \sqrt{m_N^2 + (1-x)^2 P_T^2} \Big) \Bigg]
\end{split}
\end{equation}

\iffalse
\begin{widetext}
\begin{equation}
  \frac{dN_{\rm Deu}}{d^2 P_T dy}\Big|_{y=0} = 
  C_D \, \int_0^{R} r_\perp dr_\perp\, M_T \,
  I_0 \left[ \frac{P_T \sinh \rho(r_\perp)}{T}\right] 
  \int\limits_0^1 dx \> \left|\phi_D(x)\right|^2  k_D(x,P_T) \, ,
  \label{deu_spec}
\end{equation}

%
Here, we use the notations

%
\begin{align}
  k_D(x,P_T) &=  
  K_1 \left[ \frac{\cosh \rho(r_\perp)}{T}\left(\sqrt{m_P^2 + x^2 P_T^2}+
  \sqrt{m_P^2 + (1-x)^2 P_T^2}\right) \right] 
\end{align}
\end{widetext}
%
\fi
Here $P_T$ and $M_T$ respectively denote the transverse momentum and transverse mass of the light nuclei and $m_{N} = 0.94$ GeV is the nucleon mass.
 For the functional form of $\phi_D(x)$, we choose the normalized function
\begin{align}
  \phi_D(x) &=\frac{\sqrt{\Gamma(4\alpha+2)}}{\Gamma(2\alpha+1)} \left[ x(1-x) \right]^\alpha \label{phiD}   
\end{align}

In Eq.~\eqref{deu_spec}, $C_D$ should be fixed by matching the yield of deuteron. In Eq.~\eqref{phiD}, $\alpha$ controls the slope of the deuteron spectra. Qualitatively, $\alpha$ dictates the width of the function $\phi_D$, which determines the probability of two nucleon coalescence for a given sharing of nucleus momentum between them through the momentum fraction $x$. In our model, $C_D$ and $\alpha$ are the parameters of the fit for the deuteron transverse momentum spectra. For the fit, the freeze-out parameters of neutrons are assumed to be same as that of protons since neutrons are unmeasured by ALICE. The assumption is valid at LHC energies where isospin symmetry is expected to hold.

\begin{figure*}[ht]
	\includegraphics[scale=0.28]{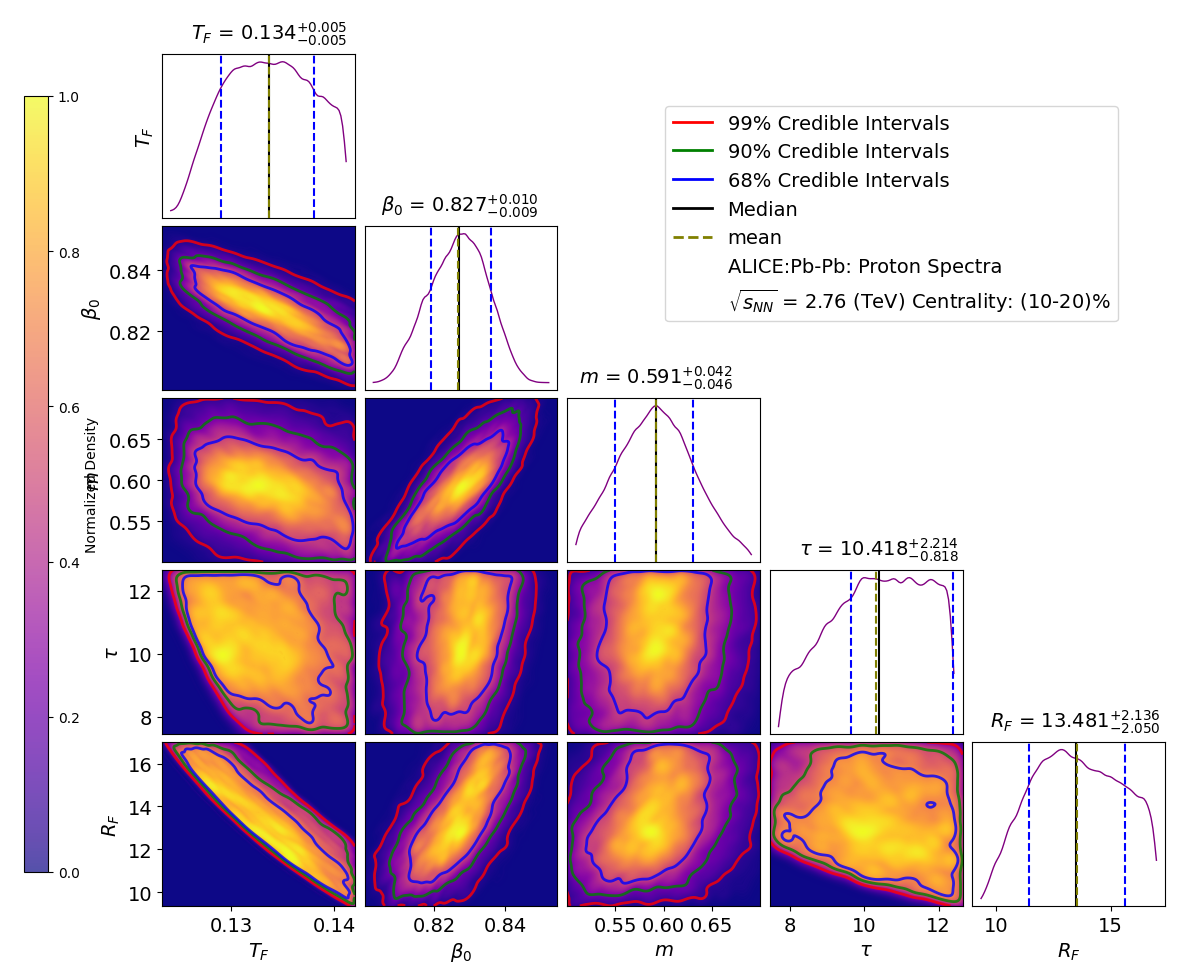}
	\includegraphics[scale=0.28]{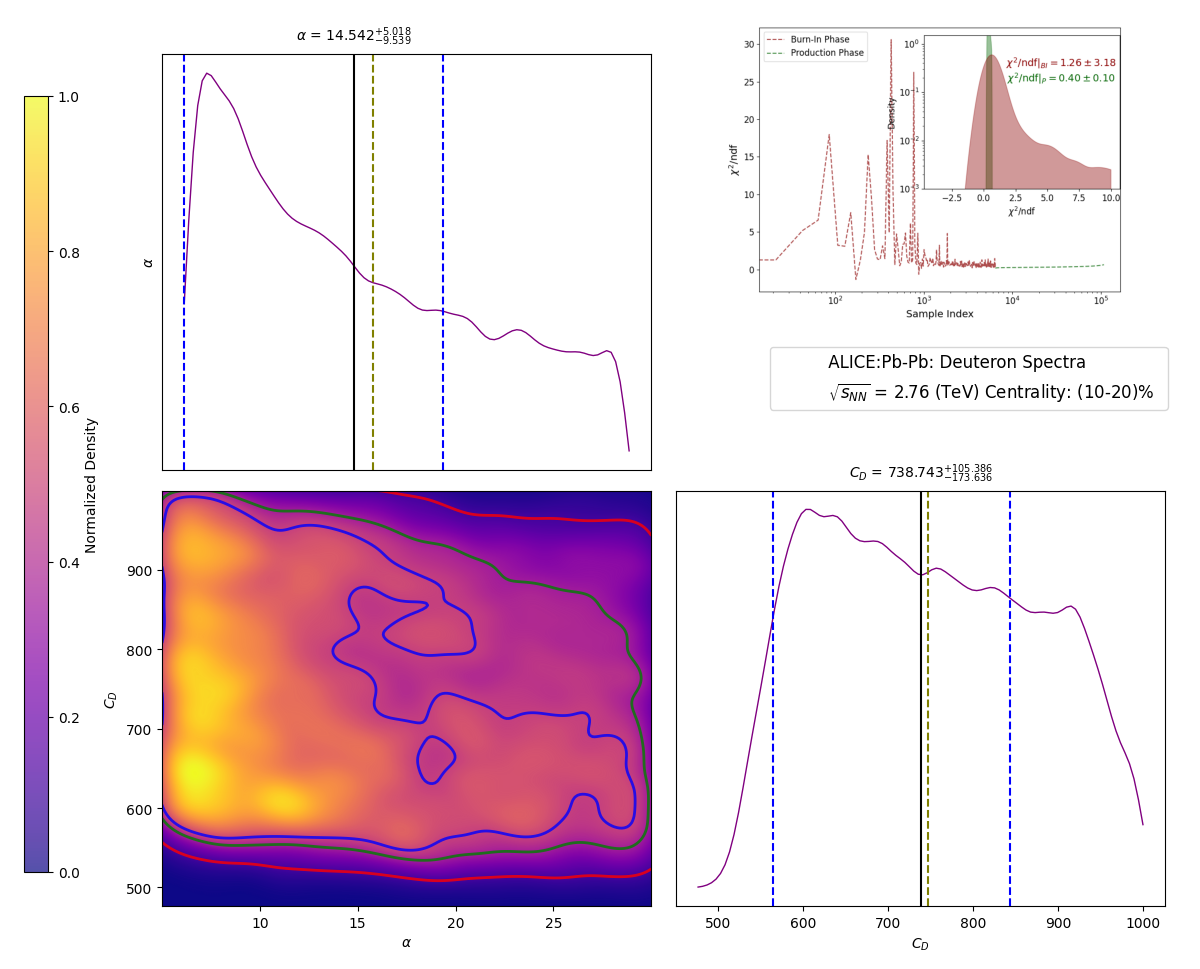}\\
	\includegraphics[scale=0.28]{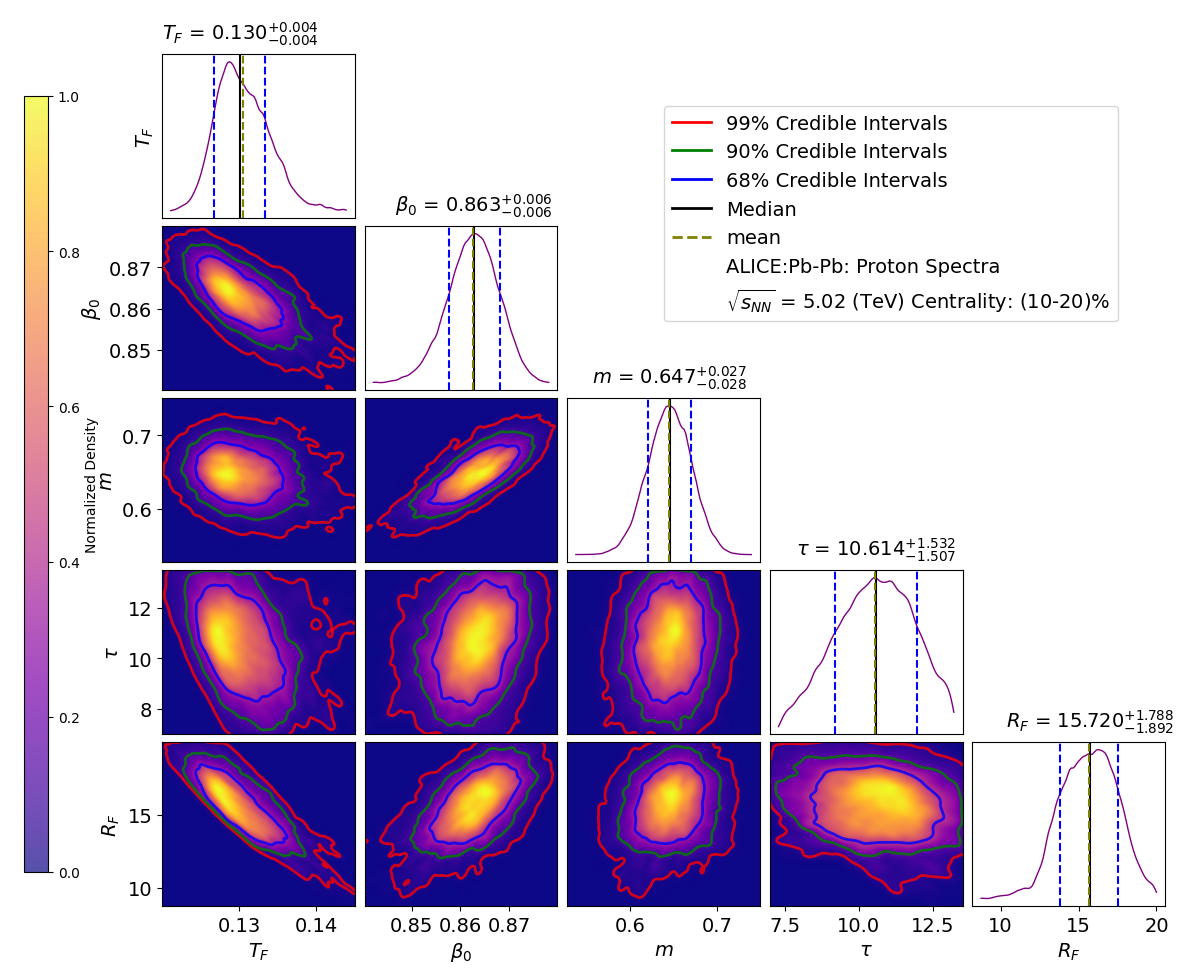} 
		\includegraphics[scale=0.28]{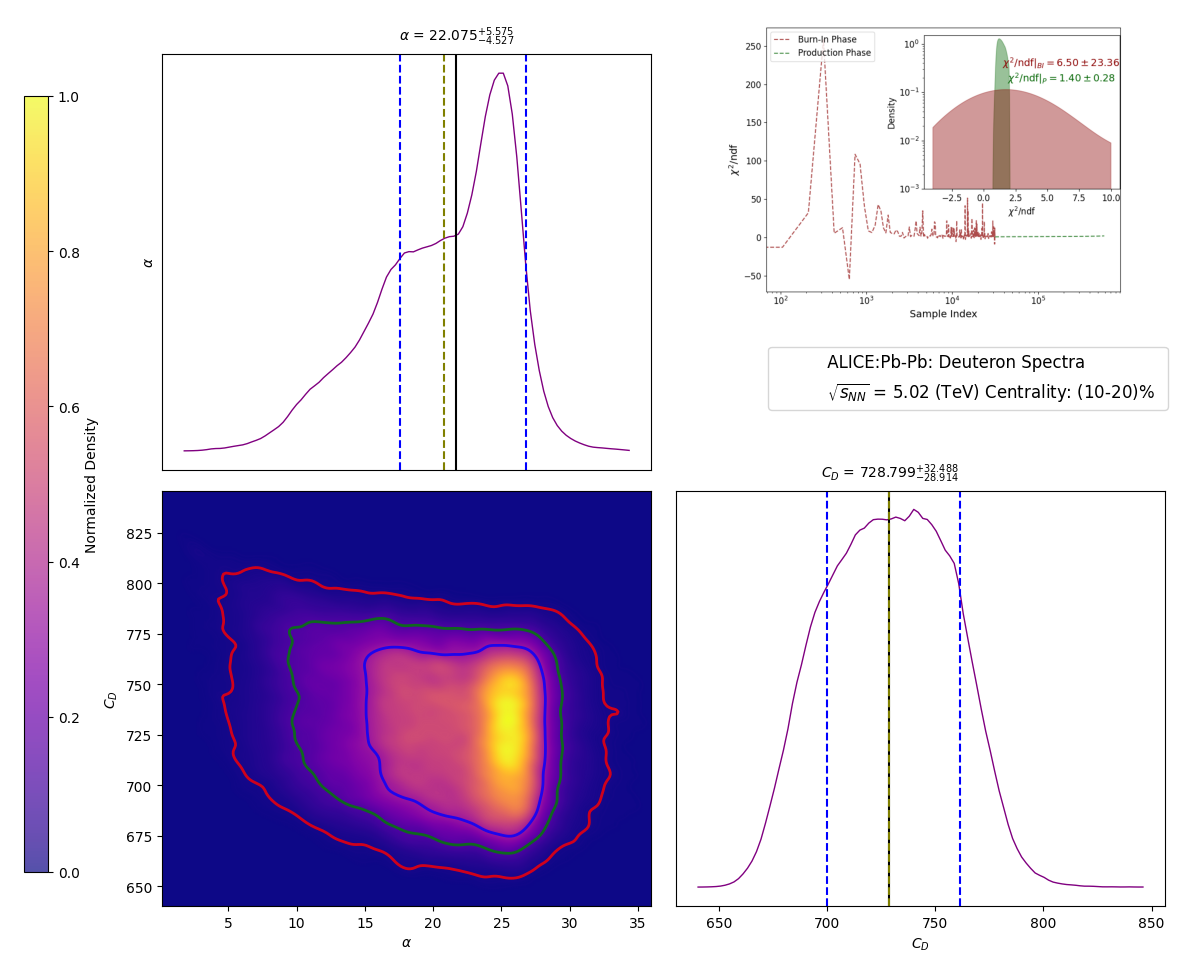} 
	\caption{Corner plot showing the Bayesian posterior distributions for proton (left) and deuteron (right) spectra at $\sqrt{s_{NN}} = 2.76$ TeV (top) and $\sqrt{s_{NN}} = 5.02$ TeV (bottom) for the centrality class (10-20)\%. The diagonal panels present histograms of marginal parameter distributions, with vertical lines marking the median, mean, and 68\% credible interval bounds. The off-diagonal panels display joint distributions between parameters, with contour lines indicating high-density regions. Color bars represent normalized density values across parameter spaces. The inset plot illustrates the $\chi^2/\mathrm{ndf}$ map for the calibration process.}
	\label{fig2}
\end{figure*}

\begin{table*}[ht]
	\centering
	\caption{Parameters for proton fit using Blast-wave model and deuteron fit using the thermo-coalescence model for Pb+Pb $\sqrt{s_{NN}}$ = 2.76 TeV  ALICE Data}
	
	\scriptsize
    \begin{tabular}{c|cccccc|ccccc}
		\toprule
		\multirow{12}{*}{\rotatebox{90}{\textbf{Proton}}} & \textbf{Parameters} & \textbf{mean} & \textbf{median} &\textbf{CI  \ (68\%)} & \textbf{CI  \ (90\%)} & \textbf{CI  \ (99\%)} & \textbf{mean} &  \textbf{median}& \textbf{CI  \ (68\%)} & \textbf{CI  \ (90\%)} & \textbf{CI  \ (99\%)} \\
		\cmidrule(lr){1-12} 
		& \multicolumn{11}{c|}{\textbf{ALICE : $\sqrt{s_{NN}}$ = 2.76 TeV (Proton and   Deuterium)}} \\
		\cmidrule(lr){1-12} 
		\multirow{8}{*}{} & \multicolumn{6}{c|}{\textbf{Centrality (0-10)\% (Interpolated)}} & \multicolumn{5}{c}{\textbf{Centrality (10-20)\%}} \\
		\midrule
		& \textbf{$T_F$}  & 135 & 135& --- & -- & -- &134& 134  &129--139  &127--142  &125--142  \\  
		& \textbf{$\beta_{0}$} & 0.83 & 0.83 & --- & --- & -- & 0.827 &0.827 & 0.818-- 0.837   & 0.812--0.841 & 0.806-- 0.846  \\
		& \textbf{m}  & 0.58 & 0.58&--- & --- & --- &0.591 &   0.591  & 0.545-- 0.633    & 0.516-- 0.657 & 0.500-- 0.687\\
		& \textbf{$\tau$}   & 10.49 & 10.49&--- & --- &---& 10.35 &  10.42  & 9.60-- 12.65  &  8.35--12.65 & 7.57--12.65    \\
		
		& \textbf{$R_F$}   & 15.58 & 15.58& --- &--- & --- &   13.53 &   13.48 & 11.43-- 15.62 & 10.94-- 16.90 & 9.87-- 17.00   \\
		
		\cmidrule(lr){1-12}

		& \multicolumn{6}{c|}{\textbf{$\bm{\chi^2/\text{ndf} = 0.41 \pm 0.17}$}
		} &\multicolumn{5}{c|}{\textbf{$\chi^2/\text{ndf} = \mathbf{0.25} \pm \mathbf{0.21}$}} \\

		\cmidrule(lr){2-12} 
		\multirow{1}{*}{\rotatebox{90}{\textbf{D}}} & \textbf{$\alpha$}   & 14.47 & 14.20&0.78--19.82 & 0.78-26.62 &0.77--29.74&15.03  & 14.94 & 0.89--20.59 & 1.07--27.45& 1.06-- 29.98  \\
		& \textbf{$C_{d}$}   & 865& 831& 686--926 &  664--1094&650--1284  & 946  & 931 & 855--996 & 807--1077 & 783--1223   \\
		\cmidrule(lr){1-12}

		\cmidrule(lr){1-12} 
		\multirow{8}{*}{} & \multicolumn{6}{c|}{\textbf{Centrality (20-40)\%}} & \multicolumn{5}{c}{\textbf{Centrality (40-60)\%}} \\
		\midrule
		& \textbf{$T_F$}  & 128 & 128& 122--135 & 120--136 & 119--141 & 122  & 121 &114--126  & 113--132 & 111--135   \\
		& \textbf{$\beta_{0}$} & 0.820 & 0.821 &0.813-0.830 & 0.804-0.836 & 0.800--0.839 & 0.808  & 0.808 & 0.798--0.820 & 0.791--0.826 & 0.783--0.834  \\
		& \textbf{m}  & 0.633 & 0.637&0.606--0.687 & 0.577--0.700 & 0.526-0.700 &0.829& 0.833 & 0.767--0.897 & 0.734--0.938  & 0.670--0.950 \\
		& \textbf{$\tau$}   & 10.23 & 10.22&8.72--11.84 & 8.35-12.64 &7.56--12.65& 9.09  & 9.01 & 6.49--10.85 & 6.28--12.34 &  5.65--12.65  \\
		& \textbf{$R_F$}   & 12.55 & 12.40& 10.13--14.70 & 9.33--15.81 & 8.73--16.95 & 11.00 & 10.89& 8.32--12.86 & 7.91--14.44 & 7.23--15.00  \\
		\cmidrule(lr){1-12} 
		
		& \multicolumn{6}{c|}{\textbf{$\bm{\chi^2/\text{ndf} = 0.29 \pm 0.22}$}
		} &\multicolumn{5}{c|}{\textbf{$\chi^2/\text{ndf} = \mathbf{0.71} \pm \mathbf{0.34}$}} \\
		\cmidrule(lr){2-12} 
		\multirow{1}{*}{\rotatebox{90}{\textbf{D}}} & \textbf{$\alpha$}   & 7.53 & 4.55&0.60--8.40 & 0.60-19.68 &0.60--28.79& 1.39  & 0.97 & 0.60--1.26 & 0.06--2.39 &  0.60--6.76  \\
		& \textbf{$C_{d}$}   & 1037& 1023& 901--1129 &  875--1221&825--1293  & 1646 &1661  & 1507--1841  &1297--1946 &  1207--1985  \\
		\cmidrule(lr){1-12}

		\cmidrule(lr){1-12} 
		\multirow{8}{*}{} & \multicolumn{6}{c|}{\textbf{Centrality (60-80)\%}} & \multicolumn{5}{c}{\textbf{---}} \\
		\midrule
		& \textbf{$T_F$}  & 118 & 118& 111--122 & 110--126 & 110--129 &  &  &  & &   \\
		& \textbf{$\beta_{0}$} & 0.772 & 0.771 &0.747-0.786 & 0.740-0.797 & 0.740--0.813 &  &  &  & &   \\
		& \textbf{m}  & 1.064 & 1.062&0.949--1.183 & 0.891--1.248 & 0.836-1.300 &&  &  &  &  \\
		& \textbf{$\tau$}   & 7.23 & 7.08&4.66--8.27 & 4.45-9.71 &4.45--10.55&  &  &  & &    \\
		& \textbf{$R_F$}   & 7.90 & 7.77& 5.65--9.06 & 5.39--10.39 & 5.01--11.42 &  &  &  & &   \\
		\cmidrule(lr){1-7}

		& \multicolumn{6}{c|}{\textbf{$\bm{\chi^2/\text{ndf} = 0.90 \pm 0.56}$}
		} &&&&&\\
		\cmidrule(lr){2-7} 
		\multirow{1}{*}{\rotatebox{90}{\textbf{D}}} & \textbf{$\alpha$}   & 0.34 & 0.31&.23--0.35 & 0.20-4.94 &0.20--5.87&  &  &  & &    \\
		& \textbf{$C_{d}$}   & 1632& 162074& 1500--1668 &  1500--1753&1500--1860  &  &  &  & &    \\
		\cmidrule(lr){1-12}

	\end{tabular}
	\label{table:alice2.76}
\end{table*}

\begin{table*}[ht]
	\centering
	\caption{Parameters for proton fit using Blast-wave model and deuteron fit using the thermo-coalescence model for Pb+Pb $\sqrt{s_{NN}}$ = 5.02 TeV  ALICE Data}
	\scriptsize
    \begin{tabular}{c|cccccc|ccccc}
		\toprule
		\multirow{12}{*}{\rotatebox{90}{\textbf{Proton}}} & \textbf{Parameters} & \textbf{mean} & \textbf{median} &\textbf{CI  \ (68\%)} & \textbf{CI  \ (90\%)} & \textbf{CI  \ (99\%)} & \textbf{mean} &  \textbf{median}& \textbf{CI  \ (68\%)} & \textbf{CI  \ (90\%)} & \textbf{CI  \ (99\%)} \\
		\cmidrule(lr){1-12} 	 	
		
		& \multicolumn{11}{c|}{\textbf{ALICE : $\sqrt{s_{NN}}$ = 5.02 TeV  (Proton and   Deuterium)}} \\
		\cmidrule(lr){1-12}

		\multirow{8}{*}{\rotatebox{90}{\textbf{ }}} & \multicolumn{6}{c|}{\textbf{Centrality (0-5)\%}} & \multicolumn{5}{c}{\textbf{Centrality (5-10)\%}} \\
		\midrule
		& \textbf{$T_F$}  & 136 & 136& 131--141 & 130--144 & 127--145 & 135& 135&129--140 &128--143& 125--145 \\
		& \textbf{$\beta_{0}$} & 0.865 & 0.865 &0.859-0.872 & 0.854-0.875 & 0.848--0.879 & 0.863 &0.864 & 0.858--0.871 & 0.853--0.874 & 0.846--0.878 \\
		& \textbf{m}  & 0.611 & 0.611&0.573--0.645 & 0.553--0.671 & 0.520-0.698 & 0.632 &0.632& 0.602--0.663 &0.584--0.683 &0.552--0.709 \\
		& \textbf{$\tau$}   & 11.08 & 11.21&10.27--13.50 & 8.86-13.49 &8.09--13.50& 10.73  & 10.89& 9.78--13.49 & 8.17--13.50 & 7.68--13.50  \\
		& \textbf{$R_F$}   & 15.66 & 15.74& 13.49--18.11 & 12.11--19.01 & 11.36--19.99 &15.18&15.35 & 13.26--17.82 & 11.46--18.45 & 10.61--19.93 \\
		\bottomrule
		& \multicolumn{6}{c|}{\textbf{$\bm{\chi^2/\text{ndf} = 1.89 \pm 0.40}$}
		} &\multicolumn{5}{c|}{\textbf{$\bm{\chi^2/\text{ndf} = 1.69 \pm 0.35}$}
		} \\
		\cmidrule(lr){2-12}

		\multirow{2}{*}{\rotatebox{90}{\textbf{D}}} & \textbf{$\alpha$}   & 26.73 & 28.68&22.83--39.84 & 13.00-39.98 &3.51--39.99& 22.75  & 23.13& 18.67--32.61 & 13.02--33.27 & 7.26--34.18  \\
		& \textbf{$C_{d}$}   & 563& 565& 540--597 &  512--608&484--634  & 618 & 619& 595--643 &  578--654&557--686  \\	
		\bottomrule

		\cmidrule(lr){1-12}
		\multirow{8}{*}{\rotatebox{90}{\textbf{Proton}}} & \multicolumn{6}{c|}{\textbf{Centrality (10-20)\%}} & \multicolumn{5}{c}{\textbf{Centrality (20-30)\%}} \\
		\midrule
		& \textbf{$T_F$}  & 130 & 130& 126--134 & 124--136 & 121--141 & 127& 126&121--129 &120--133& 120--137 \\
		& \textbf{$\beta_{0}$} & 0.863 & 0.863 &0.857-0.869 & 0.854-0.873 & 0.848--0.877 & 0.858 &0.859 & 0.853--0.865 & 0.848--0.868 & 0.841--0.872 \\
		& \textbf{m}  & 0.646 & 0.647&0.619--0.674 & 0.601--0.691 & 0.574-0.713 & 0.677 &0.678& 0.650--0.707 &0.631--0.728 &0.597--0.752 \\
		& \textbf{$\tau$}   & 10.56 & 10.61&9.11--12.15 & 8.26-13.0 4&7.36--13.49& 10.50  & 10.51& 8.49--12.45 & 7.63--13.36 & 7.14--13.98  \\
		& \textbf{$R_F$}   & 15.67 & 15.72& 13.83--17.51 & 12.84--18.53 & 11.18--19.99 &15.08&15.20 & 13.38--17.78 & 11.63--18.29 & 10.11--19.42 \\
		
		\bottomrule
		& \multicolumn{6}{c|}{\textbf{$\chi^2/\text{ndf} = \mathbf{1.41} \pm \mathbf{0.37}$}
		} &\multicolumn{5}{c|}{\textbf{$\chi^2/\text{ndf} = \mathbf{1.02} \pm \mathbf{0.28}$}} \\
		\cmidrule(lr){2-12} 
		\multirow{1}{*}{\rotatebox{90}{\textbf{D}}} & \textbf{$\alpha$}   & 22.06 &  23.07&18.55-- 28.65 & 13.06--29.77 &7.33--32.63&  21.67 &22.28&17.06--29.77 & 11.91--31.34 &6.11--33.43 \\
		& \textbf{$C_{d}$}   & 728  &   729 &  700--773  &   682--774& 664--792 & 921 & 908& 854--947 &  826--987&807--1404  \\
		\bottomrule
		
		\cmidrule(lr){1-12}  
		
		\multirow{8}{*}{\rotatebox{90}{\textbf{Proton}}} & \multicolumn{6}{c|}{\textbf{Centrality (30-40)\%}} & \multicolumn{5}{c}{\textbf{Centrality (40-50)\%}} \\
		\midrule
		& \textbf{$T_F$}  & 125 & 124& 120--128 & 119--131 & 117--135 & 125& 125&119--130 &118--133& 116--135 \\
		& \textbf{$\beta_{0}$} & 0.853 & 0.853 &0.848-0.860 & 0.843-0.864 & 0.836--0.867 & 0.841 &0.814 & 0.834--0.849 & 0.829--0.852 & 0.823--0.858 \\
		& \textbf{m}  & 0.751 & 0.752&0.719--0.783 & 0.697--0.800 & 0.678-0.828 & 0.831 &0.832& 0.801--0.870 &0.776--0.887 &0.746--0.900 \\
		& \textbf{$\tau$}   & 9.86 & 9.79&8.16--10.62 & 8.00-11.46 &8.00--11.96&9.00&8.94 & 6.81--10.44 & 6.20--11.52 & 6.00--12.35 \\
		& \textbf{$R_F$}   & 13.75 & 13.76& 11.98--16.08 & 10.64--16.59 & 10.00--17.22 &11.75&11.60 & 9.38--13.49 & 9.23--15.35 & 8.24--15.50 \\
		\bottomrule

		& \multicolumn{6}{c|}{\textbf{$\chi^2/\text{ndf} = \mathbf{1.03} \pm \mathbf{0.27}$}
		} &\multicolumn{5}{c|}{\textbf{$\chi^2/\text{ndf} = \mathbf{3.52} \pm \mathbf{0.28}$}} \\
		\cmidrule(lr){2-12} 
		\multirow{2}{*}{\rotatebox{90}{\textbf{D}}} & \textbf{$\alpha$}   & 19.24 & 19.34&12.95--28.70 & 8.09--30.41 &4.73--33.01&  7.80 &5.07&1.51--8.10 & 1.37--18.55 &1.36--28.49  \\
		& \textbf{$C_{d}$}   & 1008 & 1014& 914.--1088 &  892--1144&867--1194  & 1207 & 1204& 1107--1314 & 1059--1177&999--1400 \\
		\cmidrule(lr){1-12}

		\cmidrule(lr){1-12}
		\multirow{8}{*}{\rotatebox{90}{\textbf{Proton}}} & \multicolumn{6}{c|}{\textbf{Centrality (50-60)\%}} & \multicolumn{5}{c}{\textbf{}} \\
		\midrule
		& \textbf{$T_F$}  & 125 & 125& 120--128 & 117--133 & 114--135 &  &  &  & &   \\
		& \textbf{$\beta_{0}$} & 0.830 & 0.831 &0.823-0.838 & 0.181-0.842 & 0.812--0.847 &  &  &  & &  \\
		& \textbf{m}  & 0.992 & 0.993&0.949-1.038 & 0.919--1.065 & 0.883-1.097 &  &  &  & &  \\
		& \textbf{$\tau$}   & 7.08 & 6.89&4.61--8.04 & 4.50-9.49 &4.50--10.38&  &  &  & & \\
		& \textbf{$R_F$}   & 10.03 & 9.92& 8.02--11.47 & 7.03--12.33 & 7.00--13.96 &  &  &  & &  \\
		\cmidrule(lr){1-7}
		& \multicolumn{6}{c|}{\textbf{$\bm{\chi^2/\text{ndf} = 3.52 \pm 0.28}$}
		} &&&&&\\
		
		\cmidrule(lr){2-7}
		\multirow{2}{*}{\rotatebox{90}{\textbf{D}}} & \textbf{$\alpha$}   & 0.51 & 0.45&0.11--0.60 & $<$0.92 & $<$1.52&  &  &  & &   \\
		& \textbf{$C_{d}$}   & 1254 & 1258& 1158--1382 &  1080--1436&962--1486  &  &  &  & &  \\
		\cmidrule(lr){1-12} 
		
	\end{tabular}
	\label{table:alice5.02}
\end{table*}

%\begin{figure*}[ht]
%\includegraphics[scale=0.29]{Figures/He3_5_02_0_5_experiment_model_comparison.png}
%\includegraphics[scale=0.29]{Figures/He3_5_02_5_10_experiment_model_comparison.png}\\
%\caption{Bayesian fit of Helium3 using combined approach (the parameters from proton blast wave fits and two additinal free parameters) in 0-5$\%$ and 5-10$\%$ central Pb-Pb collisions at $\sqrt{s_{\rm NN}}=$ 5.02 TeV.}
%\label{fig3}
%\end{figure*}

%\begin{figure*}
%\includegraphics[scale=0.3]{Figures/Deuteron_fit2.jpeg}
%\caption{Comparison of Bayesian fit of deuterons using combined approach and fitting deuterons with only parameter B.} %The parameters obtained from proton fit are used to describe the deuteron spectra in addition to two more parameters.}
%
%\label{fig4}
%\end{figure*}

\section{Results and Discussions}
 \label{results and discussions}
To investigate the production mechanism of light nuclei in heavy-ion collisions, we adopt a Bayesian inference framework to extract the best-fit parameters of our hybrid model by analyzing the transverse momentum ($p_T$) spectra of protons and deuterons measured by the ALICE Collaboration~\cite{ALICE:2015wav,ALICE:2022veq}. Bayesian inference provides a statistically rigorous approach that incorporates prior knowledge along with experimental data to estimate the full posterior distribution of model parameters. Unlike frequentist methods, which often yield point estimates and confidence intervals, Bayesian inference offers a complete probabilistic characterization of parameter uncertainties, including credible intervals — ranges that directly represent the probability of a parameter lying within a given interval under the model and prior assumptions. Moreover, one of the major strengths of Bayesian inference lies in its ability to handle models with many free parameters while naturally capturing and reporting parameter correlations and degeneracies. This approach not only provides a rigorous quantification of uncertainties but also reveals intricate inter-dependencies among parameters, which are essential for interpreting complex physical models.

Due to the high computational cost of repeated evaluations of the full thermo-coalescence model across a multi-dimensional parameter space, we employ Gaussian Process Regression (GPR) emulators as surrogates \cite{Rasmussen2006,Bernhard:2018hnz}. These emulators approximate the model outputs with high fidelity while enabling efficient sampling of the posterior via Markov Chain Monte Carlo (MCMC) methods. The radial basis function (RBF) kernel used for GPR, is defined as:
\begin{equation}
k(\mathbf{x}, \mathbf{x}') = \sigma_f^2 \exp\left( -\frac{||\mathbf{x} - \mathbf{x}'||^2}{2\ell^2} \right),
\label{eq:GPR_kernal}
\end{equation}
where \( \mathbf{x} \) and \( \mathbf{x}' \) are input vectors in the parameter space, \( \ell \) is the characteristic length scale, and \( \sigma_f^2 \) is the signal variance. These are treated as hyperparameters of the model. The hyperparameters were optimized during training using a set of 300 points sampled from the multidimensional parameter space. Although the underlying model is noise-free, a small noise term was added to the RBF kernel for numerical stability during training.

Before proceeding with the inference, we validate the emulator by comparing its predictions to explicit model calculations for randomly selected 25 parameter sets(independent from the training set). As evident from Fig.~\ref{fig:GPR_Validation}, the emulator accurately reproduces the deuteron and proton $p_T$ spectra across various centralities at both $\sqrt{s_{\rm NN}} = 2.76$ and 5.02 TeV Pb-Pb collisions measured by ALICE Collaboration at LHC. The close alignment of the predicted yields with the identity line ($y = x$) confirms the reliability of the emulator.

To further understand the influence of different parameters on the model outputs, we perform a Sobol sensitivity analysis \cite{Saltelli2002}. Fig.~\ref{fig2:Sobol_index} shows the total Sobol indices ($S_T$) for the deuteron and proton spectra across $p_T$ bins covering all centrality classes at both energies. Each panel reveals how sensitive the spectra are to variations in specific model parameters, with color intensities indicating the strength of sensitivity. For deuterons, the analysis focuses on coalescence parameters ($\alpha$, $C_D$), while for protons, the relevant thermal parameters are considered. This analysis helps identify the most influential parameters in different regions of the spectra, thereby informing the interpretation of the Bayesian posteriors and guiding model refinement. It is interesting to observe that the centrality-dependent sensitivity of the parameters \( \alpha \) and \( C_D \) differs at two collision energies.
%between the two collision energies, \( \sqrt{s_{NN}} = 2.76 \)~TeV and \( \sqrt{s_{NN}} = 5.02 \)~TeV. 

Having established confidence in the emulator and identified the dominant parameters, we employ the Markov Chain Monte Carlo (MCMC) sampling method to calibrate the model parameters. We discard the initial $10^5$ samples as burn-in and generate an ensemble of $10^6$ production samples, which is subsequently used for all statistical analyses. In this analysis, we do not consider correlations among the experimentally measured data points, as experimental collaborations typically do not provide the correlation structure for systematic uncertainties. Consequently, for the simplicity the experimental covariance matrix is taken to be diagonal, with each diagonal element incorporating the total uncertainty, obtained by summing the statistical and systematic errors in quadrature. The mathematical formulation and details of the Bayesian analysis procedure can be found in~\cite{Bernhard:2018hnz}, while the analysis code used in this work has been independently developed and is available at~\cite{HepBayes}.

%\sout{Having established confidence in the emulator and identified the dominant parameters, we now present the model fits to the experimental data using Eq.~\eqref{therm}. }

The fits for the proton spectra are shown in Fig.~\ref{fig3:exp_model_com_proton}. The parameters that are kept free in the fit are, kinetic freeze-out temperature ($T_{F}$), maximum transverse flow velocity ($\beta_{0}$), exponent of flow profile (m), radius ($R_{F}$) and proper freeze-out time ($\tau$). The extracted values of these fit parameters are shown in the Table~\ref{table:alice2.76} and Table~\ref{table:alice5.02} for  Pb-Pb collisions at 2.76 TeV and 5.02 TeV respectively.  At $\sqrt{s_{NN}}= 2.76$ TeV, the data available for the most central collisions correspond to 0-5$\%$ for protons, and 0-10$\%$ for deuterons. We have interpolated the centrality dependence of the freeze-out parameters to get the freeze-out parameter values for protons in the 0-10$\%$ centrality interval. For all other centrality classes, the data for protons and deuterons are in the same interval. $T_{F}$ shows maximum of $6\%$ uncertainty with $90\%$ credible interval throughout whole centrality range, whereas, $\beta_{740}$ and $m$ show $<2\%$ and $<10\%$ within same credible interval, respectively. Moreover, comparable values of $\tau$ and $R_{F}$ are observed at most central collisions at $\sqrt{s_{\rm NN}}=$ 2.76 and 5.02 TeV. In the case of $\sqrt{s_{\rm NN}}=$ 5.02 TeV, the values of these parameters varies between 11 $fm$ to 7 $fm$ and 15 $fm$ to 10 $fm$ for $\tau$ and $R_{F}$ , respectively, with decreasing trend as a function of centrality. The larger values of freeze-out radius ($R_{F}$) relative to the freeze-out time ($\tau$) are a consequence of the accelerated expansion of the produced fireball dictated by the transverse velocity profile, Eq.~\eqref{betaT}.
One may note that, at $\sqrt{s_{NN}} = 5.02~\mathrm{TeV}$, measured data in the most peripheral centrality bin was excluded from the analysis due to large experimental uncertainties. These uncertainties resulted in extremely broad posterior distributions, making it difficult to properly constrain the corresponding model parameters within the Bayesian framework.

We move forward to use the extracted blast-wave fit parameters from proton spectra to describe the deuteron spectra. The obtained blast-wave parameters are fixed as an input to fit the deuteron spectra using two nucleon coalescence. The two coalescence parameters $C_{D}$ and $\alpha$ are kept free in fit to the deuteron spectra. The blast wave  parameters for protons and  $C_{D}$ and $\alpha$ from deuteron fit, using our model are shown in the Table~\ref{table:alice2.76} and Table~\ref{table:alice5.02} for Pb-Pb collisions at $\sqrt{s_{NN}}= 2.76$ TeV and $\sqrt{s_{NN}}= 5.02$ TeV respectively. The corresponding model fits to the deuteron $p_{T}$ spectra for these two collision energies, at different centralities are displayed in Fig.~\ref{fig1}. The model describes the experimental data remarkably well, validating this hybrid approach of light nuclei production. At $\sqrt{s_{\rm NN}}=$ 5.02 TeV, the $\alpha$ parameter is highest at the most central collisions and decreases as the collision becomes more non-central.
To visualize the outcome of the Bayesian inference and better understand the uncertainties and correlations among parameters, we present the corner plot in Fig.~\ref{fig2}. It shows the posterior distributions of the fit parameters for both proton (left) and deuteron (right) spectra at $\sqrt{s_{\rm NN}}= 2.76$ TeV (top row) and 5.02 TeV (bottom row) for the 10–20$\%$ centrality bin. The diagonal panels display the marginalized one-dimensional posterior distributions for each parameter, with the median, mean, and 68$\%$ credible intervals indicated by vertical lines. The off-diagonal panels show the two-dimensional joint posterior distributions between parameter pairs, highlighting any correlations through density contours. The color gradients reflect the normalized density across the parameter space. An inset panel in each plot displays the $\chi^2/\text{ndf}$ evaluation for both burn-in and production phases, providing a diagnostic for goodness-of-fit. The $\chi^2$ is defined as
\[
\chi^2 = (\mathbf{y}^{\text{exp}} - \mathbf{y}^{\text{model}})^\top \Sigma^{-1} (\mathbf{y}^{\text{exp}} - \mathbf{y}^{\text{model}}),
\]
with \( \Sigma \) denoting the total covariance matrix that incorporates uncertainties from both the model predictions and the experimental measurements. 
Together, these corner plots offer detailed insights into parameter uncertainties, degeneracies, and the robustness of the model calibration.
The posterior correlations among the blast-wave parameters reveal how different aspects of the fireball's freeze-out dynamics interplay. A strong correlation between $\beta_0$ (maximum transverse flow) and $m$ (flow profile exponent) arises because both influence the spectral slope: steeper flow profiles can mimic stronger flow, highlighting a degeneracy in extracting collective expansion features from transverse momentum spectra alone. Similarly, the mild positive correlation between $R_f$ and $\tau$ reflects the geometric growth of the system over time—larger systems tend to freeze out later. An observed anti-correlation between $T_f$ and the flow parameters indicates a compensatory relationship between thermal smearing and radial flow in shaping the $p_T$ spectra. In the case of deuterons, the absence of a notable correlation between the coalescence parameters $\alpha$ and $C_D$ carries clear physical significance. The parameter $\alpha$ governs the internal momentum correlation between constituent nucleons through the shape of the deuteron wavefunction, while $C_D$ encodes the spatial overlap probability relevant for coalescence. This statistical independence indicates that the spectral shape and the overall yield of deuterons are influenced by fundamentally different physical mechanisms. Such decoupling underscores the capability of the hybrid framework to disentangle the microscopic structure of the coalescence kernel from the macroscopic freeze-out geometry.

\begin{figure}[ht!]
\centering
\includegraphics[width=\linewidth]{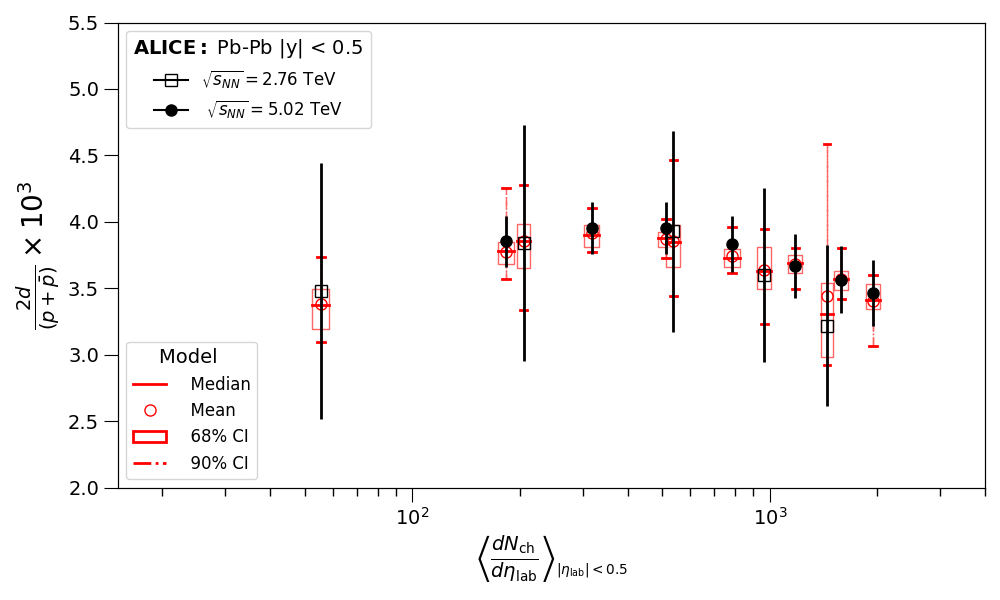}
\caption{Ratio of deuteron to proton yield as a function of charged particle multiplicity.}
\label{yield_ratio}
\end{figure}

We move on by computing the integrated yields of protons and deuterons across different centrality classes by numerically integrating the corresponding spectra. This allows us to extract the total particle yields within the measured $p_{T}$ range for each centrality bin. We then evaluate the deuteron-to-proton yield ratio, defined as $2d/(p + \bar{p})$, which serves as an important observable for probing the underlying production mechanism of light nuclei. The calculated yield ratios are compared with experimental results reported by the ALICE Collaboration, as shown in Fig.~\ref{yield_ratio}. Our model successfully reproduces the centrality dependence of the yield ratio within the 68$\%$ credible intervals derived from the Bayesian analysis, demonstrating the robustness of our hybrid framework in describing both the spectral shapes and integrated yields. This agreement further strengthens the validity of the coalescence mechanism employed in our approach.

\begin{figure}[H] % requires \usepackage{float}
\centering
\begin{minipage}{0.48\textwidth}
    \includegraphics[width=\linewidth]{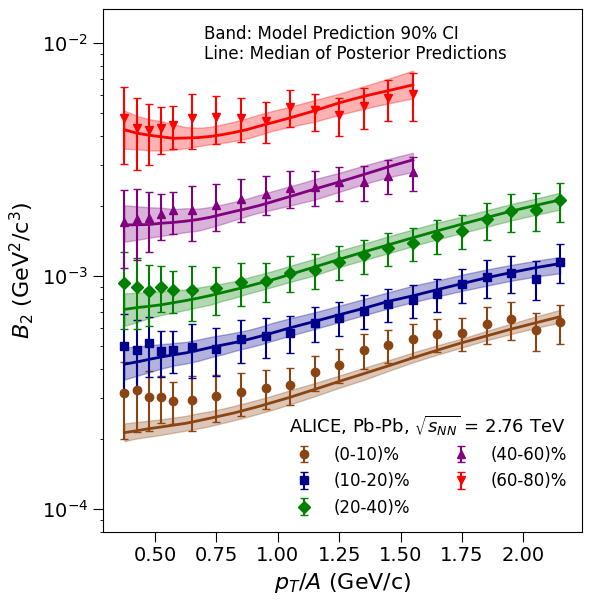}
\end{minipage}
\hfill
\begin{minipage}{0.48\textwidth}
    \includegraphics[width=\linewidth]{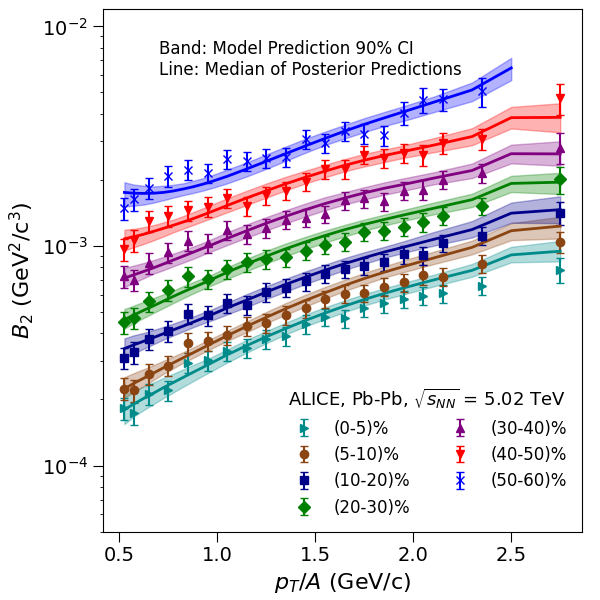}
\end{minipage}
\caption{Comparison of the coalescence parameter $B_2$ between $\sqrt{s_{NN}}= 2.76$ TeV and $5.02$ TeV Pb--Pb collisions across different centrality classes.}
\label{fig:B2_comparison}
\end{figure}

\begin{figure*}[ht]
\centering
\includegraphics[scale=0.5]{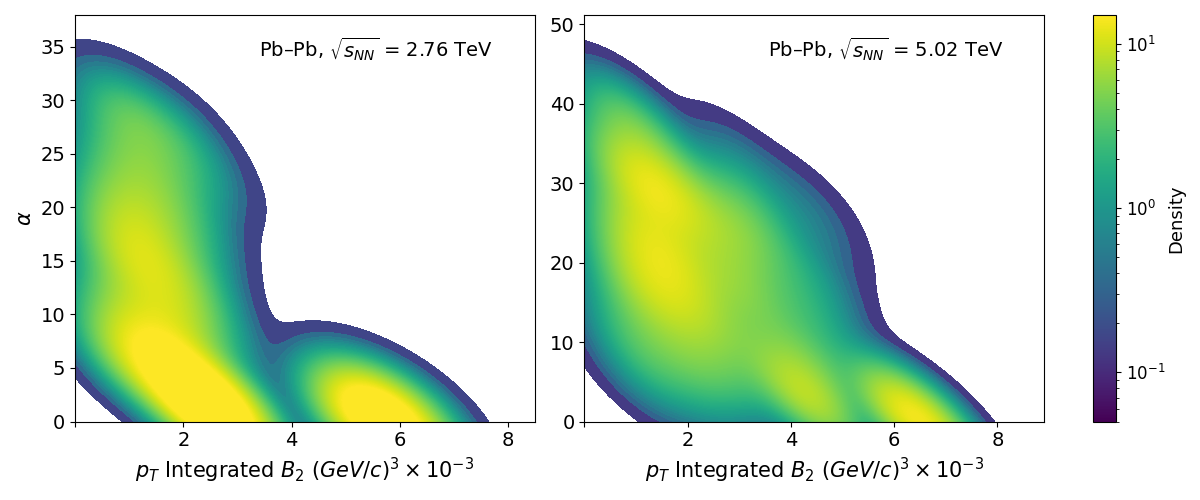}
  \caption{
Correlation plot of the coalescence parameter \( B_2 \) (integrated over \( p_T \)) and the parameter \( \alpha \) for Pb--Pb collisions at \(\sqrt{s_{NN}} = 2.76\)~TeV (left panel) and \(5.02\)~TeV (right panel). A common color scale indicates the relative density in both panels.
}
    \label{B2_alpha_corelation}
\end{figure*}

Traditionally the coalescence model used for describing light nuclei production is based on coalescence factor $B_2$ (for deuterons) defined as\cite{ALICE:2022veq}:

\begin{equation}\label{eq:B2}
E\,\frac{dN_D}{d^3P} = B_2 \left( E\,\frac{dN_p}{d^3p}\right) \left( E\,\frac{dN_n}{d^3p}\right) \simeq B_2 \left( E\,\frac{dN_p}{d^3p}\right)^2,
\end{equation}

Here, $P$ is the deuteron momentum and $p$ is the proton momentum. 
%Since $B_{2}$ exhibit $p_{\rm T}$ dependence, the spectra is not fitted well.
The coalescence parameter $B_2$ represents the probability of coalescence of a neutron and a proton to form a deuteron. As the value of $B_2$ can be calculated once the spectra of deuteron and proton have been determined, we have attempted to estimate it using our models that describe the proton and deuteron spectra. The obtained results for Pb-Pb collisions at $\sqrt{s_{NN}}= 2.76$ and   $\sqrt{s_{NN}}= 5.02$ TeV are shown in Fig.~\ref{fig:B2_comparison}  and match well with the measurements from ALICE Collaboration \cite{ALICE:2015wav,ALICE:2022veq}. The ALICE Collaboration results have been obtained by using a simple blast wave inspired fit to the deuteron spectra, and then taking the spectral ratio following Eq.(\ref{eq:B2}). For a given value of $p_{T}/A$ the gradual decrease of $B_{2}$ from peripheral to central collisions, can be attributed to the increasing size of the fireball, which makes it more difficult for a $pn$ pair to recombine into a deuteron. In the same centrality bin, the rise in $B_{2}$  with $p_{T}/A$ is due to the decrease of the available volume with increasing momentum of the particle~\cite{ALICE:2015wav}. It may be interesting to note that the present model quantitatively reproduces the centrality and $p{T}$ dependence of $B_{2}$ without assuming any explicit space-momentum correlation in its theoretical construct. On the other hand, previous studies~\cite{Wang:2023rpd, Wang:2024jpe} have reported that recombination models with coordinate momentum factorization cannot reproduce the observed trend in $B_{2}$. Coalescence models including space-momentum correlation can better reproduce the data.

Finally before we close, note that our model parameter $\alpha$ shows a decreasing trend from central to peripheral collisions. It would be thus interesting to study the correlation between parameter $\alpha$ of our hybrid approach and recombination probability $B_2$. We observe a clear anti-correlation between the two as shown in Fig~\ref{B2_alpha_corelation}. Relatively larger values of  parameter $\alpha$ in central collisions, narrows the deuteron internal wavefunction to allow coalescence for only a narrow range of the momentum fraction variable $x$ in Eq.(\ref{phiD}). On the other hand, as we go towards peripheral collisions, the wavefunction spreads out and allows for coalescence for a wider range of momentum fractions. Thus larger values of $\alpha$ correspond to narrower wavefunctions, which reduce the overlap between nucleons and consequently suppress the coalescence probability, leading to smaller values of $B_2$.

\section{Conclusions}
\label{conclusions}
In this work, a combined approach of thermal and coalescence production, so-called, "thermo-coalescence model" has been employed to describe the light nuclei production in ultra-relativistic nuclear collisions at LHC. The underlying basis of the model is that the nucleons are assumed to be thermally produced and equilibrated, subsequently, undergo coalescence to eventually form light nuclei. In this paper, we only focus on deuterons to validate applicability of this model. To this end, we analyzed $p_{\rm T}-$spectra of protons and deuterons for various available centralities and centre-of-mass energies in Pb-Pb collisions at LHC. To model the phase space distribution of the primordial nucleons we carried out individual Bayesian fits to the $p_{T}$ spectra of prompt protons at both energies and for various collision centrality, following a boost invariant blast wave model, with five free parameters. The extracted fit parameters were fixed an inputs to deuteron spectra with two free parameters. The model has described the ALICE measurements of deuteron transverse spectra reasonably well, including the $p_{T}$ and centrality dependence of the coalescence parameter $B_{2}$. The $\alpha$ parameter shows a centrality dependence with highest value for most central collisions. The model has been further extended to estimate the protons and deuteron yields and described the experimentally measured deuteron to proton ratios at various charge particle multiplicities within the uncertainties. It would be further interesting to extend this model to study the elliptic flow coefficient of deuterons and differential and inclusive productions of three body bound states like helium-3 and tritons as well as their sensitivity to the space-momentum correlation of nucleons at the kinetic freeze-out. We aim to report these studies in a future communication.

\end{document}